\documentclass[10pt]{iopart}
\pdfoutput=1

\usepackage{etoolbox}
\usepackage{hyperref}

\usepackage{lipsum}

\makeatletter
\def\@mkboth#1#2{}
\newlength\appendixwidth
\preto\appendix{\addtocontents{toc}{\protect\patchl@section}}
\newcommand{\patchl@section}{%
  \settowidth{\appendixwidth}{\textbf{Appendix }}%
  \addtolength{\appendixwidth}{1.5em}%
  \patchcmd{\l@section}{1.5em}{\appendixwidth}{}{\ddt}%
}
\makeatother

\expandafter\let\csname equation*\endcsname=\relax
\expandafter\let\csname endequation*\endcsname=\relax
\usepackage{color}
\usepackage{amsmath}
\usepackage{amssymb}
\usepackage{enumerate}
\usepackage{graphicx}
\usepackage{mathbbol}
\usepackage{amsfonts}
\usepackage{url}

\newcommand{\x}{{\bf x}}
\newcommand{\p}{{\bf p}}

\newcommand{\cyan}{\textcolor{cyan}}

\newcommand{\bea}{\begin{eqnarray}}
\newcommand{\eea}{\end{eqnarray}}
\newcommand{\beq}{\begin{equation}}
\newcommand{\eeq}{\end{equation}}

\def\XXint#1#2#3{{\setbox0=\hbox{$#1{#2#3}{\int}$}
 \vcenter{\hbox{$#2#3$}}\kern-.5\wd0}}

\usepackage[utf8]{inputenc}
\usepackage{amsmath}
\usepackage{hyperref}
\usepackage{graphicx}
\usepackage{amsfonts}
\usepackage{amsthm}
\usepackage{cases}
\usepackage{bm}
\usepackage{amssymb}

\usepackage{color}
\definecolor{Blue}{rgb}{0.00, 0.00, 1.00}
\definecolor{Red}{rgb}{1.00, 0.00, 0.00}

\hypersetup{
    colorlinks=true,       
    linkcolor=red,          
    citecolor=blue,        
    filecolor=magenta,      
    urlcolor=cyan           
}

\newcommand{\be}{\begin{equation}}
\newcommand{\ee}{\end{equation}}

\newcommand{\beqn}{\begin{eqnarray}}
\newcommand{\eeqn}{\end{eqnarray}}

\DeclareMathOperator{\Ai}{Ai}

\DeclareMathOperator{\J}{J}

\def\q{\frac{\hbar^2}{2m}}

\makeatletter
\renewcommand\@appendixstar{\@@par
 \ifnumbysec 
 \@addtoreset{table}{section}
 \@addtoreset{figure}{section}\fi
 \setcounter{section}{0}
 \setcounter{subsection}{0}
 \setcounter{subsubsection}{0}
 \setcounter{equation}{0}
 \setcounter{figure}{0}
 \setcounter{table}{0}
 \def\thesection{\Alph{section}} 
 \def\theequation{\ifnumbysec
      \Alph{section}.\arabic{equation}\else
      \Alph{section}\arabic{equation}\fi}
 \def\thetable{\ifnumbysec
      \Alph{section}\arabic{table}\else
      A\arabic{table}\fi}
 \def\thefigure{\ifnumbysec
      \Alph{section}\arabic{figure}\else
      A\arabic{figure}\fi}}
\makeatother

\begin{document}
\title[Noninteracting fermions in a trap and RMT]{Noninteracting fermions in a trap and random matrix theory}

\author{David S. Dean}
\address{Univ. Bordeaux and CNRS, Laboratoire Ondes et Mati\`ere d'Aquitaine (LOMA), UMR 5798, F-33400 Talence, France}

\author{Pierre Le Doussal}
\address{CNRS-Laboratoire de Physique Th\'eorique de l'Ecole Normale Sup\'erieure, 24 rue Lhomond, 75231 Paris Cedex, France}

\author{Satya N. Majumdar}
\address{LPTMS, CNRS, Univ. Paris-Sud, Universit\'e Paris-Saclay, 91405 Orsay, France}

\author{Gr\'egory Schehr}
\address{LPTMS, CNRS, Univ. Paris-Sud, Universit\'e Paris-Saclay, 91405 Orsay, France}

\begin{abstract}
We review recent advances in the theory of trapped fermions using techniques borrowed from random matrix theory (RMT) and, more generally, from the theory of determinantal point processes. In the presence of a trap, and in the limit of a large number of fermions $N \gg 1$, the spatial density exhibits an edge, beyond which it vanishes. While the spatial
correlations far from the edge, i.~e. close to the center of the trap, are well described by standard many-body techniques, such as the local density approximation (LDA), these methods fail to describe the fluctuations close to the edge of the Fermi gas, where the density is very small and the fluctuations are thus enhanced. It turns out that RMT and determinantal point processes offer a powerful toolbox to study these edge properties in great detail. Here we discuss the principal edge universality classes, that have been recently identified using these modern tools. In dimension $d=1$ and at zero temperature $T=0$, these universality classes are in one-to-one correspondence with the standard universality classes found in the classical unitary random matrix ensembles: soft edge (described by the ``Airy kernel'') and hard edge (described by the ``Bessel kernel'') universality classes. We further discuss extensions of these results to higher dimensions $d\geq 2$ and to finite temperature. Finally, we discuss correlations in the phase space, i.e., in the space of positions and momenta, characterized by the so called Wigner function. 
\end{abstract}

\maketitle

\tableofcontents

\newpage

\section{Introduction}

Over the past few decades, there have been spectacular experimental developments in manipulating cold atoms (bosons or fermions) \cite{BDZ08, GPS08} that
have led to a number of Nobel prizes. These developments allow one to probe quantum many-body physics, both for interacting and noninteracting systems. In these systems the nature of the interaction can be tuned experimentally and the effective interaction can actually be removed. However, even noninteracting bosons and fermions display interesting collective many-body effects emerging purely from the quantum statistics \cite{Mahan, Castin, Castin2}. For noninteracting fermions, which we focus on here, the Pauli exclusion principle induces
highly non-trivial spatial (and temporal) correlations between the particles. Remarkably, the recent development of 
Fermi quantum microscopes \cite{Cheuk:2015,Haller:2015,Parsons:2015} provides a direct access to these spatial correlations, via a direct 
in situ imaging of the individual fermions, with a resolution comparable to the inter-particle spacing. The theoretical understanding of these spatio-temporal correlations in noninteracting fermions is therefore an outstanding and challenging problem.

\begin{figure}
\centering
\includegraphics[width = \linewidth]{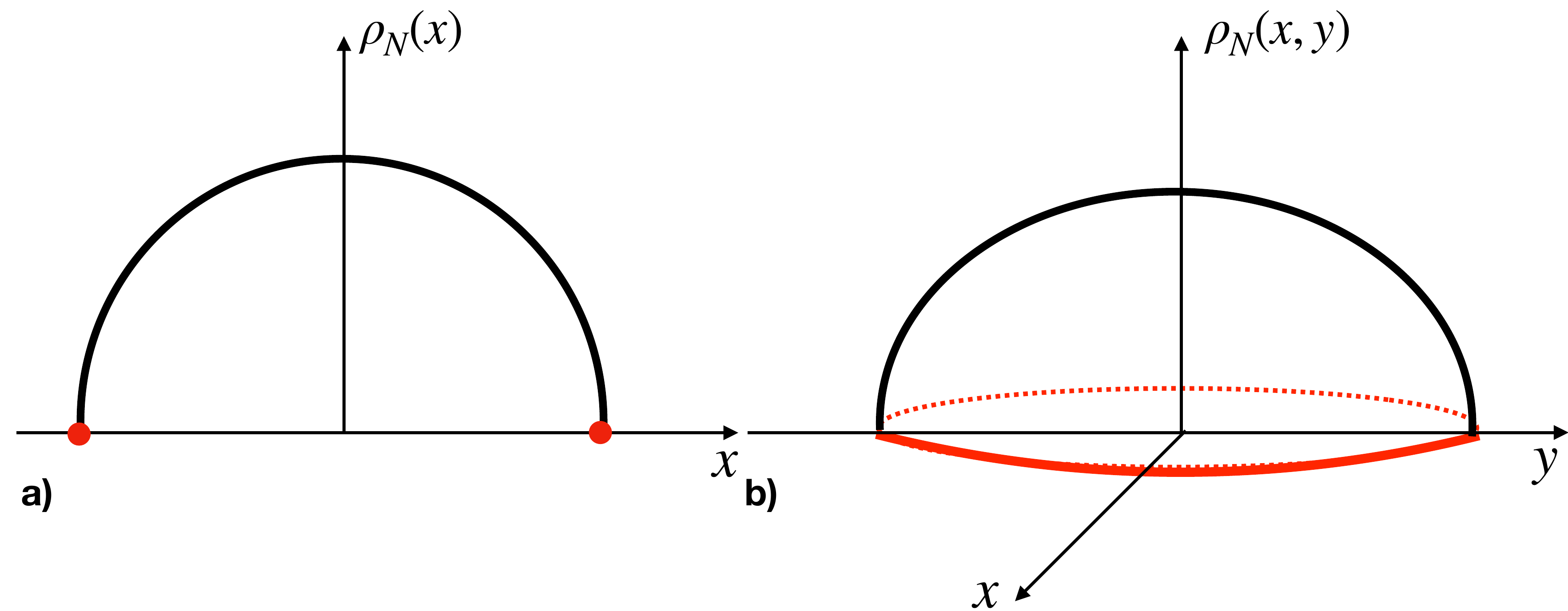}
\caption{Sketch of typical density profiles for non-interacting trapped fermions, a) in $d=1$ and b) in $d=2$. In both cases, it exhibits an edge (in red) beyond which the (scaled) density vanishes in the limit $N \to \infty$: a) in $1d$ the ``edge'' consists of two points while in $d=2$ the edge is a circle (for spherically symmetric potential).}\label{Fig_edge}
\end{figure}

In contrast to classical systems, quantum systems display non-trivial spatial fluctuations even at zero temperature ($T=0$) due to the zero-point motion of the particles. These purely quantum fluctuations, in combination with the quantum statistics of particles (Bose-Einstein or Fermi-Dirac), give rise to non-trivial spatial correlations. The presence of a confining trap also affects these spatial correlations in a non-trivial way and this is our main object of interest here. Indeed, the confining trap breaks the translational invariance of the system. The physics in the bulk near the trap center (where the fermions do not feel the curvature of the 
confining trap) can be understood using the traditional theories of quantum many-body systems such as the local density 
approximation (LDA)~\cite{Castin,butts}. However, away from the trap center, the fermions start feeling the curvature induced by the confining 
trap. As a result the average density profile of the fermions vanishes beyond a certain distance from the trap center---thus 
creating a {\em sharp edge}, see Fig. \ref{Fig_edge}. Near this edge, the density is small (there are few fermions) and consequently, quantum and thermal fluctuations 
play a more dominant role than in the bulk. The importance of these fluctuations means that traditional theories such as LDA 
break down in this edge region. Indeed, this was pointed out by Kohn and Mattson that {\em the uniform electron gas, the traditional starting point for density-based many-body theories of inhomogeneous systems, is inappropriate near electronic edges} \cite{Kohn}. One thus needs new methods to describe this edge physics. In this review we will demonstrate a connection to Random Matrix Theory (RMT) which can thus be exploited to provide  precise and powerful tools to
address the edge physics (see Fig. \ref{Fig_V_RMT}).  The methods we discuss will be used to derive the average density profile for the free fermionic system,  but also the two point kernel from which all
statistics and correlation functions can be inferred.

In a series of recent studies we have shown how the techniques from RMT can be exploited to make precise predictions for the spatial correlations between noninteracting fermions near the edge \cite{marino_prl,us_finiteT,DPMS:2015,fermions_review}. In one dimension and at zero temperature, the joint distribution of the positions of the fermions in a trap (that characterizes the purely quantum fluctuations) can be mapped, for certain types of traps, to the joint distribution of eigenvalues of an appropriate classical random matrix ensemble (see Fig. \ref{Fig_V_RMT}). For example, the harmonic potential corresponds to the Gaussian Unitary Ensemble (GUE), the hard box potential corresponds to the Jacobi Unitary Ensemble (JUE) and the potential $V(x) = A\,x^2 + B/x^2$ ($x>0$) corresponds to the Laguerre Unitary Ensemble (LUE). The fact that these ensembles are all unitary reflects the fact that the fluctuations are quantum in nature, as we will see below. In dimensions $d>1$ or at temperature $T>0$, these direct connections to RMT ensembles no longer hold. However, the underlying structure of spatial correlations is still described by a determinantal point process (DPP) which is completely characterized by a temperature and dimension dependent kernel. In this short review, we briefly discuss some of these developments involving RMT and describe how it leads to precise predictions for the spatial correlations in this trapped Fermi gas, both in the bulk as well as at the edges. In the bulk, our results recover in a controlled way the results of the LDA, which is extensively used in the atomic physics literature. However, at the edge, RMT techniques lead to new results which can not be obtained using the semi-classical (LDA) approximation.  

The paper is organised as follows. We start in Section \ref{1drm} by explaining the exact correspondence  between the position of $N$ spin-less trapped fermions at zero temperature in one dimension, with a number of confining potentials, and the eigenvalues of a number of unitary Gaussian random matrix ensembles. 
These exact correspondences allow a number of results from random matrix theory to be directly transposed to the context of trapped fermions. In Section \ref{ds} we describe the determinantal structure of the statistics of the trapped fermion problem. In particular, we show how all correlation functions can be expressed in terms of a kernel and how this kernel behaves in the limit of a large $N$. 
In particular, we show how the statistics are strongly modified at the edge of the Fermi gas, where the effects of quantum fluctuations are much more important than in the bulk. In section (\ref{ldim}) we  consider trapped fermions at zero temperature in higher dimensions $d\geq 2$. Although the direct link with RMT no longer holds in this case, these systems still possess the determinantal structure exhibited by those in one dimension. We present results for the behaviour of the average density and kernel as a function of spatial dimension, both in the bulk and at the edge. In Section \ref{temp} we examine what happens at non-zero temperature. There we show that in the canonical ensemble, i.e. for fixed particle number, the determinantal structure is lost. However it is recovered if one passes to the grand canonical ensemble. By exploiting this we can obtain the bulk and edge properties, in a well defined low temperature regime, by using the equivalence between the canonical and grand canonical ensembles in the thermodynamic limit. In Section \ref{Sec:phase_sp}, we consider the correlations in the phase space, i.e. in position and momentum space $(x,p)$, characterized by  
the so-called Wigner function, which also exhibits an edge in the $(x,p)$ plane. Focusing on the edge in momentum space, we also discuss 
some recently discovered connections with multi-critical matrix models. Finally, we conclude in Section \ref{Sec:conclusion}.

\section{$1d$ noninteracting trapped fermions at $T=0$ and random matrix ensembles}\label{1drm}

We consider $N$ spinless noninteracting fermions in a one-dimensional trapping potential $V(x)$. The system is thus described by the $N$-body Hamiltonian $\hat{\cal H}_N = \sum_{j=1}^N \hat h_j$ where $\hat h_j = \hat h(\hat x_j, \hat p_j)$ is a single-particle Hamiltonian of the form
\bea\label{def_H1d}
\hat h = \frac{\hat p^2}{2\,m} + V(\hat x) \;.
\eea
Let us denote by $\phi_l(x)$ the $l$-th single-particle eigenfunction ($l = 1,2, \cdots$) with eigenvalue $\epsilon_l$, i.e.,
\bea \label{eigen_f}
\hat h \, \phi_l(x) = \epsilon_l \phi_l(x) \;.
\eea
The ground-state of the $N$-body system corresponds to filling up the $N$ first single-particle energy levels with
one fermion per level (as dictated by the Pauli exclusion principle). Correspondingly, the $N$-body ground-state wave-function 
is given by the Slater determinant
\bea\label{Slater_1d}
\Psi_0(x_1, \cdots, x_N) = \frac{1}{\sqrt{N!}} \, \det_{1\leq \,j,\,l\, \leq N}\phi_{l}(x_j)  \;,
\eea
with the associated energy $E_0 = \sum_{l=1}^N \epsilon_l$. The quantum probability density function (PDF) is then given by
\bea\label{jpdf}
P_{\rm joint}(x_1, \cdots, x_N) = |\Psi_0(x_1, \cdots, x_N)|^2 = \frac{1}{N!} \left|\det_{1\leq \,j,\,l\, \leq N} \phi_{l}(x_j) \right|^2 \;.
\eea
This joint PDF is normalised  and encodes the quantum fluctuations of the Fermi gas. For an arbitrary potential $V(x)$ it is hard to
solve this Schr\"odinger~equation (\ref{eigen_f}) and evaluate explicitly the Slater determinant in (\ref{Slater_1d}). However, for a few specific potentials $V(x)$ the Slater determinant can be computed  as we show below. 

\subsection{Harmonic potential and the GUE}

We consider first the harmonic trap $V(x) = \frac{1}{2} m\,\omega^2 x^2$. In this case, the single-particle eigenfunctions $\phi_k(x)$ are given by
 \bea\label{def_Hermite}
\phi_k(x) = \left[ \frac{\alpha}{\sqrt{\pi} 2^k k!}\right]^{1/2} \, e^{-\frac{\alpha^2\,x^2}{2}} H_k(\alpha \,x) \;,
\eea
where $k=0, 1, \cdots$ (note that here, and what follows, the index $k$ starts at $0$ while the generic index $l$ in Eq. (\ref{eigen_f}) starts at 1),  $H_k(z)$ is the $k$-th Hermite polynomial of degree $k$ and $\alpha = \sqrt{m \omega/\hbar}$ is the characteristic length scale of the trap. The associated single-particle energy levels are given by $\epsilon_k = (k+1/2)\, \hbar \omega$. Here, to construct the Slater determinant, we take the first $N$ energy levels labelled by $k=0, \cdots, N-1$. In the Slater determinant, the Gaussian factors come out of the determinant, leaving us to compute the determinant of a matrix consisting of Hermite polynomials. The Hermite polynomials $H_0(z), H_1(z), \cdots, H_{N-1}(z)$ provide a basis for polynomials of degree $N-1$ and by  manipulating the rows and columns, the determinant can  be reduced to a Vandermonde determinant. Hence, we can evaluate the Slater determinant explicitly to obtain
\bea\label{j_PDF_GUE}
P_{\rm joint}(x_1, \cdots, x_N) = \frac{1}{z^{\rm GUE}_N} e^{-\alpha^2 \sum_{i=1}^N x_i^2}\,\prod_{i<j}(x_i-x_j)^2 \;,
\eea 
where $z^{\rm GUE}_N$ is a normalisation constant. We identify immediately that, up to a trivial rescaling factor $\alpha$, this is precisely the joint distribution of the eigenvalues of a $N \times N$ GUE matrix of RMT \cite{mehta,forrester}. Clearly the Vandermonde square term $\prod_{i<j}(x_i-x_j)^2$ provides an effective repulsion between any pair of fermions coming purely from the Pauli exclusion principle. Thus even though the fermions are noninteracting to start with, their quantum statistics provides an effective pairwise repulsion. Note that in the context of GUE eigenvalues, the Vandermonde square term has a purely mathematical origin, coming from the Jacobian of the transformation from matrix entries to eigenvalues and eigenvectors \cite{mehta}. Finally, we notice that, since in quantum mechanics, the probability density is always the square of the modulus of the wave function, the power of the Vandermonde term is naturally 2, and hence the corresponding random matrix ensemble is necessarily a unitary ensemble.

\begin{figure}
\centering
\includegraphics[width = \linewidth,angle = 0]{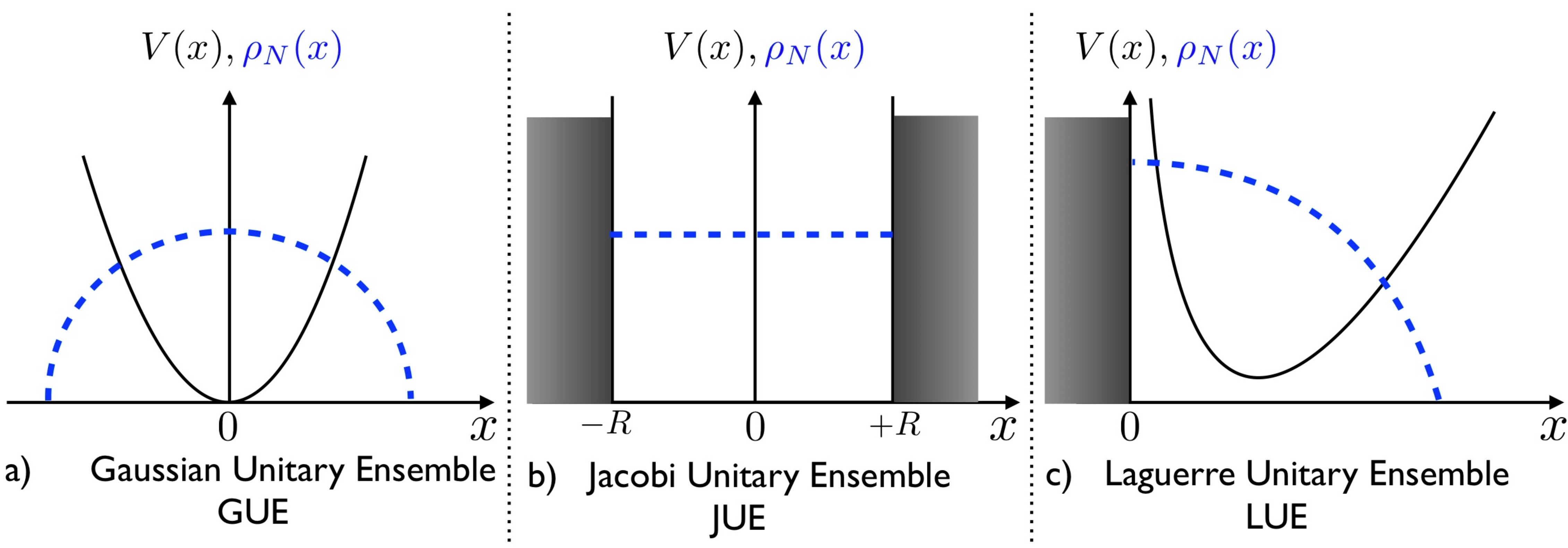}
\caption{Quantum potentials $V(x)$ (black solid line) and associated bulk fermion density $\rho_N(x)$ (blue dotted line) corresponding to three different unitary ensembles of RMT discussed here: a) the harmonic potential $V(x) = m\omega^2\, x^2/2$, for which the limiting density is the Wigner semi-circle (\ref{wigner}), corresponds to the GUE (\ref{j_PDF_GUE}), b) the hard box potential on $[-R,+R]$, for which the limiting density is uniform (\ref{unif_density}), corresponds to the JUE as in (\ref{JUE_gen}) with parameters $a=b=1/2$ (more general JUE ensembles correspond to potentials of the form (\ref{pot_JUE})) and c) the potential $V(x) = A \, x^2 + B/x^2$ on $(0, + \infty)$ (\ref{def_inv_sq}) for which the density is a ``half'' semi-circle (\ref{density_inv_sq}), corresponds to the LUE (\ref{joint_LUEsq}).} \label{Fig_V_RMT}
\end{figure}

Before we discuss other potentials, it is interesting to point out one immediate consequence of the one-to-one mapping between the positions of the fermions in a $1d$ harmonic trap at zero temperature and the eigenvalues of GUE. In the RMT literature there has been a tremendous recent interest in the distribution of the largest eigenvalue $\lambda_{\max}$ (for short reviews see \cite{maj,MS_thirdorder}). When appropriately centered and scaled, the limiting distribution of $\lambda_{\max}$ is  the celebrated Tracy-Widom (TW) GUE law~\cite{TW94}. This TW distribution has since appeared in a wide variety of, apparently unconnected,  problems \cite{baik,johann,poli,growth,SS10,CLR10,DOT10,ACQ11,sequence,dots,FMS11,Lie12,biroli} and has also been measured in experiments \cite{takeuchi,davidson}, albeit somewhat indirectly. From the above mapping (\ref{j_PDF_GUE}), we see that the position $x_{\max}$ of the rightmost fermion at $T=0$ corresponds to $\lambda_{\max}$ and, hence, the quantum fluctuations of $x_{\max}$, appropriately centered and scaled, is also described by the TW-GUE law. This provides a possibility to directly measure the TW-GUE distribution in trapped fermion systems \cite{us_finiteT}.

\subsection{Hard box potential and the JUE}

Let us now consider the case of a hard box potential $V(x)$ of the form
\be\label{hb_pot_1d}
V(x)=\begin{cases}
        &\displaystyle 0\;,\;\;|x|\leq R\\
        & \displaystyle\infty\;,\;\;|x|>R\;.
       \end{cases}
\ee
In this case, the single-particle Schr\"odinger equation (\ref{eigen_f}) can be solved exactly with eigenfunctions and energies given by 
\be\label{hb_1d_wf}
\phi_l(x)=\sin\left(\frac{l\pi}{2R}(x+R)\right)\;\;{\rm and}\;\;\epsilon_l=\q k_l^2=\frac{\hbar^2\pi^2}{8 m R^2}l^2\;,
\ee
for $l=1,2,\cdots$. We set in the following $R=1$, which amounts to rescaling all positions by $R$. The $N$-body ground state wave function is given by the 
Slater determinant constructed from the single-particle eigenfunctions in Eq. \eqref{hb_1d_wf},
\be\label{Slater_det}
\Psi_0({x}_1,\cdots,{x}_N)=\frac{1}{\sqrt{N!}}\det_{1\leq \, j, \,l \,\leq N}\phi_l(x_j)=\frac{1}{\sqrt{N!}}\det_{1\leq \,j, \,l \,\leq N}\sin\left(\frac{l \pi}{2}(x_j+1)\right)\;.
\ee
This Slater determinant can be written in a more convenient way by using the identity $\sin(nx)=\sin(x)U_{n-1}(\cos(x))$ where $U_n(t)$ is the Chebychev polynomial of second kind of degree $n$. By rearrangements of rows and columns, the joint quantum PDF of the positions in Eq. (\ref{jpdf}) reads~\cite{FFGW03,Cunden1D,LLMS17,LLMS18}
\be\label{joint_PDF_hb_1d}
P_{\rm joint}({x}_1,\cdots,{x}_N)=\frac{1}{z^{\rm JUE}_N}\prod_{l=1}^N \cos^2\left(\frac{\pi x_l}{2}\right)\prod_{i<j}^N \left|\sin\left(\frac{\pi x_i}{2}\right)-\sin\left(\frac{\pi x_j}{2}\right)\right|^2\;,
\ee
where $z^{\rm JUE}_N$ is a normalisation constant. Introducing the new variables $u_i=(1+\sin(\pi x_i/2))/2$, the joint PDF of $u_1,\cdots,u_N$ can be worked out from \eqref{joint_PDF_hb_1d}. 
It coincides with the joint PDF of the eigenvalues of a matrix belonging to the JUE~\cite{mehta,forrester,Cunden1D,FFGW03}
\be\label{JUE}
P_{\rm joint}(u_1,\cdots,u_N)=\frac{1}{\tilde z_N^{\rm JUE}}\prod_{l=1}^N \sqrt{u_l(1-u_l)}\prod_{i<j}^N \left|u_i-u_j\right|^2\;, \;\; u_i \in [0,1]\;.
\ee
This is of course a special case of a more general Jacobi ensemble \cite{forrester}
\bea\label{JUE_gen}
P_{\rm joint}(u_1,\cdots,u_N)\propto \prod_{k=1}^N u_l^a\,(1-u_l)^b\prod_{i<j}^N \left|u_i-u_j\right|^2\;, \;\; u_i \in [0,1]\;,
\eea
parametrised by two real numbers $a>-1$ and $b>-1$. The joint PDF for the hard box potential  in Eq. (\ref{JUE}) corresponds to $a=b=1/2$. It is natural to ask the question  if there exist quantum potentials that correspond to the general JUE with arbitrary parameters $a$ and $b$. Indeed, it was shown recently \cite{LLMS18} that a potential of the type
\bea\label{pot_JUE}
V(x) = \frac{a^2-\frac{1}{4}}{8\sin^2(\frac{x}{2})} + \frac{b^2-\frac{1}{4}}{8\cos^2(\frac{x}{2})} \;, \;\; x \in [0,\pi]
\eea
generates a joint PDF of the form in (\ref{JUE_gen}) with arbitrary $a$ and $b$.

\subsection{The potential $V(x) = A\,x^2 + B/x^2$, with $x>0$, and the LUE}

Here we consider a potential of the form $V(x) = A\,x^2 + B/x^2$ which we conveniently parametrise as follows \cite{NM_interface}
\bea\label{def_inv_sq}
V(x) = 
\begin{cases}
&\dfrac{b^2}{2} x^2 + \dfrac{\alpha(\alpha-1)}{2x^2} \;,\; x>0  \;, \\
& \\
& + \infty \;, \; \hspace*{2cm} x \leq 0 \;,
\end{cases}
\eea
with $b>0$ and $\alpha > 1$. For convenience, we set here $\hbar = 1$ as well as the mass $m=1$. For a potential of this form~(\ref{def_inv_sq}), the Schr\"odinger equation (\ref{eigen_f}), together with the boundary condition $\phi_k(0) = 0$ (since we impose a hard wall at $x=0$), can be solved exactly. The single-particle eigenfunctions $\phi_l(x)$ and associated energies $\epsilon_l$ are given~by
\bea\label{ev_laguerre}
\phi_k(x) = c_k\, e^{-\frac{b}{2}x^2} x^\alpha {\cal L}_k^{(\alpha-\frac{1}{2})}(b\,x^2) \;, \; \epsilon_k=b\left(2k+ \alpha + \frac{1}{2}\right) \;,
\eea
where $k = 0,1, \cdots$ is a non-negative integer, $c_k$ is a normalisation constant and ${\cal L}_k^{(\alpha-\frac{1}{2})}(z)$ is a generalized Laguerre polynomial of degree $k$. Constructing the Slater determinant out of the first $N$ states, one gets (again using the fact that the determinant of orthogonal polynomials, in this case generalized Laguerre polynomials, reduces to a Vandermonde form)
\bea\label{joint_LUEsq}
P_{\rm joint}(x_1, \cdots, x_N) \propto e^{-b \sum_{i=1}^N x_i^2} \prod_{l=1}^N x_l^{2 \alpha} \prod_{i<j} (x_i^2 - x_j^2)^2 \;.
\eea
Making further the change of variables $y_i = x_i^2$, the joint PDF of the $y_i$'s reads \cite{NM_interface}
\bea 
\hspace*{-1cm}P_{\rm joint}(y_1, \cdots, y_N) = \frac{1}{z_N^{\rm LUE}} e^{-b \sum_{i=1}^N y_i} \prod_{l=1}^N y_l^{\alpha-\frac{1}{2}} \prod_{i < j} (y_i - y_j)^2 \;, \;\; y_i \geq 0 \;.
\eea
This corresponds to the joint PDF of the eigenvalues of a Wishart-Laguerre unitary ensemble (LUE) of random matrices \cite{mehta,forrester}.

We end this section with the following remark. In the original fermion problem, there is an external quantum potential $V(x)$. We have shown that, for some choices of this $V(x)$, the Slater determinant square can be interpreted as the joint PDF of the eigenvalues of a corresponding unitarily invariant random matrix ensemble. It is natural to ask the reverse question. Suppose we start with a unitarily invariant random matrix ensemble, where the
entries of an $N \times N$ complex matrix $X$ are distributed as ${\rm Pr}(X) \propto e^{-{\rm Tr}V_M(X)}$ where $V_M(X)$ is typically a polynomial matrix potential. Given $V_M(X)$, one can ask if there is a fermion problem with a suitable potential $V(x)$ whose Slater determinant square  
 would correspond to the joint PDF of this RMT ensemble. For GUE (corresponding to $V_M(X) = X^2$), we have seen above that $V(x)$ is also 
 a harmonic potential. However, for a general $V_M(X)$, it is not clear that there is an underlying fermion problem with a suitable quantum potential $V(x)$.

\section{Determinantal structure of the spatial correlations in $d=1$ and $T=0$}\label{ds}

For noninteracting fermions in an arbitrary potential $V(x)$, all the information about the (quantum) spatial fluctuations are contained in the joint PDF in Eq. (\ref{jpdf}). Of special interest are the $n$-point spatial correlation functions $R_n(x_1, \cdots, x_n)$, with $1 \leq n \leq N$, which are given by the different marginals of the full joint PDF, i.e.,~\cite{mehta,forrester}
\be \label{def_correl} 
R_n(x_1, \cdots, x_n) =  \frac{N!}{(N-n)!} \int d{x}_{n+1} \cdots \int d {x}_N \,P_{\rm joint}(x_1, \cdots, x_n, x_{n+1}, \cdots, x_N) \;,
\ee
where the integrals over the positions $x_i$'s run over their full domain of definition (and which thus depends on the quantum potential). In particular, for $n=1$ 
\be \label{def_density}
R_1(x) = N \, \int dx_2 \cdots \int d x_N P_{\rm joint}(x, x_2, \cdots, x_N) \;,
\ee
which is directly related to the average density of fermions in the ground-state via
\be \label{rel_R1_density}
R_1(x) = N\,\rho_N(x) \;, \;\;\; \rho_N(x) = \frac{1}{N}\left\langle \sum_{i=1}^N \delta(x-x_i) \right\rangle_0  \;,
\ee
where $\langle \cdots \rangle_0$ denotes an average in the ground state $\Psi_0(x_1, \cdots, x_N)$ (\ref{jpdf}). Note that this density $\rho_N(x)$ is normalized to unity, and not to the total number of fermions.

To perform the multiple integrals in Eq. (\ref{def_correl}) or (\ref{def_density}), it is convenient 
to rewrite the joint PDF in (\ref{jpdf}) as
\bea\label{jpdf_2}
P_{\rm joint}(x_1, \cdots, x_N) = \frac{1}{N!} \det_{1\leq \,j,\,l\, \leq N} \phi^*_{l}(x_j)  \det_{1\leq \,j,\,l\, \leq N} \phi_{l}(x_j) \;.
\eea
Using the property $\det (A^T) \det (B) = \det (AB)$, the product of two determinants  in (\ref{jpdf_2}) can be written as a single determinant 
\bea \label{jpdf_3}
P_{\rm joint}(x_1, \cdots, x_N) = \frac{1}{N!} \det_{1\leq \,j,\,l\, \leq N} K_\mu(x_j, x_l) \;,
\eea
where we have introduced the kernel $K_\mu(x,y)$ defined by
\bea\label{def_kernel_1d}
K_\mu(x,y) = \sum_{l=1}^N \theta(\mu-\epsilon_l) \phi^*_l(x) \phi_l(y) \;,
\eea
where $\theta(z)$ is the Heaviside theta function, i.e. $\theta(z) = 1$ if $z > 0$ and $\theta(z) = 0$ if $z<0$, and $\mu$ is the Fermi energy (here this is simply the energy of the last occupied level, i.e., $\mu = \epsilon_N$). Exploiting  the ortho-normality of the single-particle eigenfunctions, i.e., $\int dx \, \phi^*_l(x) \phi_{l'}(x) = \delta_{l,l'}$, it is easy to check that the kernel $K_{\mu}(x,y)$ in (\ref{def_kernel_1d}) is self-reproducible, i.e., it satisfies the important property
\bea \label{self_reprod_1d}
\int dy \, K_\mu(x,y) K_\mu(y,z) = K_\mu(x,z) \;.
\eea    
This property plays an important role because it implies that the $n$-point correlation function $R_n(x_1, \cdots, x_n)$ can be written as an $n \times n$ determinant \cite{mehta,forrester} 
\bea\label{determinantal_structure}
R_n(x_1, \cdots, x_n) = \det_{1 \leq \, j,\, l \, \leq n} K_\mu(x_j,x_l) \;,
\eea
for any $1 \leq n \leq N$. In particular, for $n=1$, this result (\ref{determinantal_structure}), together with the relation in (\ref{rel_R1_density}), implies
\bea\label{coinciding}
\rho_N(x) = \frac{1}{N} K_\mu(x,x) = \frac{1}{N} \sum_{l=1}^N |\phi_l(x)|^2 \;.
\eea
This property (\ref{determinantal_structure}) establishes that the positions of $N$ noninteracting fermions trapped in an arbitrary potential $V(x)$ constitute a determinantal point process \cite{J05, Bo11} with a kernel $K_\mu(x,y)$ given by Eq. (\ref{def_kernel_1d}).  

Before analysing the large $N$ behaviour of the kernel, we present a few important and useful properties of determinantal processes. Let us first consider the number of fermions $N_{\cal I}$ within an interval ${\cal I} = [a,b]$: we would like to describe the statistics of $N_{\cal I}$ in the ground state $\Psi_0(x_1, \cdots, x_N)$ given in Eq. (\ref{Slater_1d}), i.e., compute the generating function $\langle z^{N_{\cal I}}\rangle_0$, from which the full counting statistics for the fermions within the interval ${\cal I}$ can be obtained. To this purpose, it is useful to introduce the indicator function $\chi_{\cal I}(x)$ defined as
\bea\label{indicator}
\chi_{\cal I}(x) = 
\begin{cases}
&1 \;, \; {\rm if} \; x \in {\cal I} \\
&0 \;, \; {\rm if} \; x \notin {\cal I} \;.
\end{cases}
\eea
Hence $N_{\cal I}$ can be written as $N_{\cal I} = \sum_{i=1}^N \chi_{\cal I}(x_i)$, which implies that $z^{N_{\cal I}} = \prod_{i=1}^N z^{\chi_{\cal I}(x_i)}$. Therefore the generating function can be written as 
\bea\label{GF1}
\langle z^{N_{\cal I}}\rangle_0 = \left \langle \prod_{i=1}^N (1-(1-z)\chi_I(x_i))\right \rangle_0 \;,
\eea
where we have used that $z^{\chi_{\cal I}(x)} = 1 - (1-z)\chi_{\cal I}(x)$ for the binary variable $\chi_{\cal I}(x)$ (\ref{indicator}). Since the $x_i$'s form a determinantal point process (\ref{determinantal_structure}), the average in the right hand side of Eq. (\ref{GF1}) can be written as~\cite{J05,Bo11}
\bea \label{FCS_fred}
\langle z^{N_{\cal I}}\rangle_0 = {\rm Det}\left({\mathbb 1} - (1-z) \chi_{\cal I} \, K_\mu \, \chi_{\cal I}\right) \;,
\eea
where ${\rm Det}$ denotes a Fredholm determinant [we recall that ${\rm Det}({\mathbb 1} - \tilde K) = \exp(-\sum_{p\geq 1} {\rm Tr}\tilde K^p/p)$], $K_\mu \equiv K_{\mu}(x,y)$ is the kernel in (\ref{def_kernel_1d}) and $\chi_{\cal I} \equiv \chi_{\cal{I}}(x)$ is the projector on the interval $\cal{I}$ (\ref{indicator}) -- and therefore $\chi_{\cal I} K_\mu \chi_{\cal I} \equiv \chi_{\cal I}(x) K_\mu(x,y) \chi_{\cal I}(y)$. From this exact formula (\ref{FCS_fred}) it is then possible to extract the cumulants of $N_{\cal I}$ and, in principle, recover the full distribution of $N_{\cal I}$. For instance the probability that there is no fermion in the interval ${\cal I}$, ${\rm Pr}.(N_{\cal I} = 0)$, is simply given by the right hand side of Eq. (\ref{FCS_fred}) evaluated at $z=0$,
\bea\label{pr_hole}
{\rm Pr}.(N_{\cal I} = 0) =  {\rm Det}\left({\mathbb 1} - \chi_{\cal I} \, K_\mu \, \chi_{\cal I}\right) \;.
\eea
Specialising this formula (\ref{pr_hole}) to the case ${\cal I} = [M, + \infty)$ yields the cumulative distribution of the position of the rightmost fermion $x_{\max}(T=0) = \max_{1\leq i \leq N} x_i$. Indeed, ${\rm Pr}.(x_{\max}(T=0) \leq M) = {\rm Pr}.(N_{[M,+\infty)} = 0)$ and therefore, from Eq. (\ref{pr_hole}), we obtain immediately
\be\label{fred_xmax}
{\rm Pr}.(x_{\max}(T=0) \leq M) = {\rm Det}\left({\mathbb 1} - \chi_{\cal I} \, K_\mu \, \chi_{\cal I}\right) \;, \; {\rm with}\;\;\; {\cal I} = [M, + \infty) \;.
\ee
These results in Eqs.  (\ref{determinantal_structure}), (\ref{FCS_fred}) and (\ref{fred_xmax}) show that a huge amount of information can be obtained from the 
kernel $K_{\mu}(x,y)$ (\ref{def_kernel_1d}), which is thus a central object. 

Of course, for finite $N$, the kernel (\ref{def_kernel_1d}) and thus these different observables (\ref{determinantal_structure}), (\ref{FCS_fred}) and (\ref{fred_xmax}) will depend on the specific form of the trapping potential $V(x)$ in (\ref{def_H1d}). But what happens in the large $N$ limit? Quite generically, in the presence of a trapping potential $V(x)$, the density $\rho_N(x)$, for $N \gg 1$, has a finite support $[-x_{\rm edge}, +x_{\rm edge}]$ (for simplicity we consider here a symmetric potential $V(x) = V(-x)$) and thus it exhibits {\it edges} at $x = \pm x_{\rm edge}$ beyond which the density vanishes. Far from the edges, in the {\it bulk}, the density can be computed using the LDA~\cite{Castin, butts}. The starting point of the LDA is a semi-classical approximation of the so-called Wigner function $W_N(x,p)$ (see Section \ref{Sec:Wigner} below), which can be interpreted as a (pseudo) single-particle probability distribution over the phase space $(x,p)$: by integrating $W_N(x,p)$ over $p$ one obtains the spatial density $\rho_N(x)$ and by integrating it over $x$ one obtains the density in momentum space [see Eq. (\ref{marginal_Wigner}) below]. At finite inverse temperature $\beta = 1/T$, the LDA approximates $W_N(x,p)$ by the Fermi-factor (up to a prefactor) corresponding to the total energy $E(x,p) = p^2/(2m)+V(x)$
\bea \label{wigner_finite_beta}
W_N(x,p) \approx \frac{1}{2 \pi \hbar} \frac{1}{e^{\beta(E(x,p) - \tilde \mu)} + 1} \;,
\eea 
with $\tilde \mu$ the finite temperature chemical potential. At $T=0$, i.e. $\beta \to \infty$,  $\tilde \mu = \mu$ and the Fermi factor in (\ref{wigner_finite_beta}) reduces to a simple Heaviside theta-function, i.e. 
\bea \label{wigner_T0}
W_N(x,p) \approx \frac{1}{2 \pi \hbar} \theta\left(\mu - \frac{p^2}{2m} - V(x)\right) \;.
\eea
By integrating (\ref{wigner_T0}) one obtains the LDA prediction for the density 
\bea\label{rho_LDA}
\rho_N(x)  = \int W_N(x,p)\, dp \approx \frac{\sqrt{2m}}{N \pi \hbar} \theta(\mu - V(x)) \left[\mu - V(x) \right]^{1/2} \;, 
\eea
which has a finite support $[-x_{\rm edge}, + x_{\rm edge}]$ where the edge is thus defined as 
\bea \label{edge}
V(x_{\rm edge}) = \mu \;,
\eea
and we recall that $\mu$ is the Fermi energy, i.e., here the last occupied single-particle energy level in the many-body ground state. Since the density has a finite support (\ref{rho_LDA}), one naturally expects that the kernel $K_\mu(x,y)$ in (\ref{def_kernel_1d}) will exhibit a different behaviour in the bulk, for $x,y$ far from the edges, and close to the edges, for $x \sim y \sim x_{\rm edge}$ (or equivalently $x \sim y \sim -x_{\rm edge}$), see Fig. \ref{Fig_length}. In the bulk, for generic $x$ and $y$ with a separation of the order of the local inter-particle distance, i.e., $|x-y| \sim 1/(N \rho_N(x))$, the kernel $K_\mu(x,y)$ takes the scaling form   
\bea \label{k_bulk}
K_\mu(x,y) \approx \frac{1}{\ell(x)} {\cal K}_{\rm Sine}\left(\frac{x-y}{\ell(x)}\right)   \;, \; \ell(x) = \frac{1}{N\pi \rho_N(x)} \;,
\eea 
where the scaling function ${\cal K}_{\rm Sine}(z)$ is universal, i.e., independent of $V(x)$ \cite{fermions_review,Eis2013}, and given by the sine-kernel
\bea\label{sine_k}
{\cal K}_{\rm Sine}(z) =  \frac{\sin z}{\pi\,z} \;,
\eea
which is well known in RMT \cite{mehta,forrester}. While this result in the bulk can also be obtained using the LDA \cite{Castin} or semi-classical approaches, these methods fail to study the large $N$ behaviour of the kernel near the edges \cite{Kohn}. It  is precisely in this region where the RMT tools are very useful. Indeed, these questions related to the edge of the spectrum of random matrices have generated a lot of interest during the last twenty years in the RMT literature \cite{TW94} (for a short review see \cite{MS_thirdorder}). In particular, it is well known in RMT that 
the different matrix ensembles corresponding to the three different fermion models mentioned above (GUE, JUE and LUE) lead to different behaviours at the edge. Therefore, below, we study the edge behaviours in the three different models separately.

\subsection{Harmonic potential (GUE): soft edge scaling and the Airy kernel}

\begin{figure}
\centering
\includegraphics[width = 0.7\linewidth]{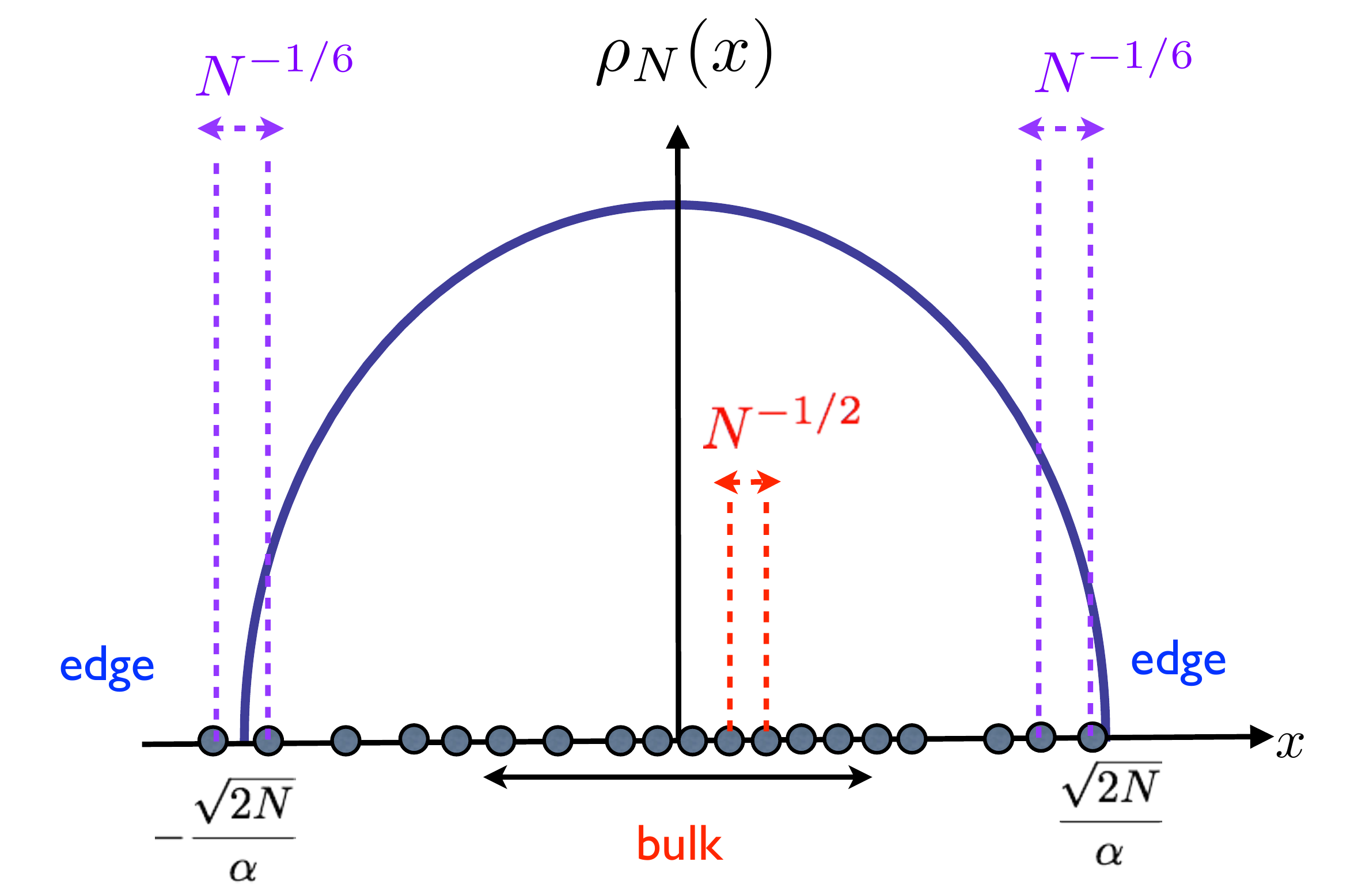}
\caption{Illustration of the different length scales both in the bulk and at the edge for the $1d$ harmonic potential $V(x) = m\omega^2 x^2/2$. The solid line represents the bulk density $\rho_N(x)$, given in this case by the Wigner semi-circle (\ref{wigner}), which has a finite support $[-x_{\rm edge}, + x_{\rm edge}]$, with $x_{\rm edge} = \sqrt{2N}/\alpha$ and $\alpha = \sqrt{m \omega/\hbar}$. In the bulk, close to the center of the trap, the typical inter-particle distance is $\ell(0) = {\cal O}(N^{-1/2})$ [see Eq.~(\ref{k_bulk})]. In contrast, the inter-particle distance at the edge, close to $\rm x_{\rm edge}$ is much larger and given by $w_N = {\cal O}(N^{-1/6})$ [see Eq.~(\ref{wN})].}\label{Fig_length}
\end{figure}

For the harmonic potential $V(x) = \frac{1}{2}m \omega^2 x^2$, the limiting density given by (\ref{rho_LDA}) is the well know Wigner semi-circle, which takes the scaling form
\begin{eqnarray}\label{wigner}
\rho_N(x) \approx 
\frac{\alpha}{\sqrt{N}} 
f_W\left(\frac{\alpha \, x}{\sqrt{N}} \right) \;, \; 
f_W(z) = \frac{1}{\pi}\sqrt{2-z^2} \;,
\end{eqnarray}
with soft edges at $x = \pm x_{\rm edge}$ with $x_{\rm edge} = \sqrt{2N}/\alpha$. Therefore, the inter-particle distance in the bulk (\ref{k_bulk}) is of order $\ell(x) = {\cal O}(N^{-1/2})$. In contrast, one expects that, near the edge, the inter-particle distance $w_N$ is much larger, as $\ell(x) \to \infty$ for $x \to \pm x_{\rm edge}$. In fact $w_N$ can be estimated by considering that the fraction of particles in the interval $[x_{\rm edge}-w_N, x_{\rm edge}]$ is of order ${\cal O}(1/N)$, i.e. 
\bea \label{estimate_evs}
\int_{x_{\rm edge}-w_N}^{x_{\rm edge}} \rho_N(x) \, dx \approx \frac{1}{N} \;.
\eea
Given the behaviour of the bulk density (\ref{wigner}) near $x=x_{\rm edge} = \sqrt{2N}/\alpha$, one obtains
\bea\label{wN}
w_N = \frac{1}{\alpha \sqrt{2}} N^{-1/6} \;,
\eea  
which is the width of the edge region (see Fig. \ref{Fig_length}), well known for GUE \cite{TW94}. Near the (soft) edge, for both $x, y \simeq x_{\rm edge}$, the kernel takes the scaling form
\bea\label{k_soft_edge}
K_\mu(x,y) \approx \frac{1}{w_N} {\cal K}_{\rm Ai}\left(\frac{x-x_{\rm edge}}{w_N}, \frac{y-x_{\rm edge}}{w_N}\right) \;,
\eea
where ${\cal K}_{\rm Ai}(z,z')$ is the Airy kernel
\be
{\cal K}_{\rm Ai}(z,z') = \frac{{\rm Ai}(z){\rm Ai}'(z')-{\rm Ai}'(z){\rm Ai}(z')}{z-z'}\, = \int_0^{+\infty} du \, {\rm Ai}(z+u) {\rm Ai}(z'+u) \,,
\label{airy_kernel.1}
\ee
where ${\rm Ai}(x)$ is the Airy function. In particular, from this result (\ref{airy_kernel.1}), together with the relation (\ref{coinciding}), we obtain the density profile near the edge. The sharp edge of the bulk density in Eq. (\ref{wigner}) is smeared out, for large but finite $N$, over a length $\sim w_N$ close to the edges $\pm x_{\rm edge}$ where it is described 
by a finite size scaling form (say close to the right edge $+x_{\rm edge}$)~\cite{BB91,For93} 
\begin{eqnarray}\label{edge_density_1d}
\rho_N(x) \approx \frac{1}{N \, w_N} 
F_{\rm Ai}\left[\frac{x - x_{\rm edge}}{w_N} \right] \;,
\end{eqnarray}
where the scaling function is given by~\cite{BB91,For93} 
\bea\label{edge_density_scaling_function}
F_{\rm Ai}(z) = [{\rm Ai}'(z)]^2 - z [{\rm Ai}(z)]^2\, \;.
\eea
The scaling function $F_{\rm Ai}(z)$ has 
the asymptotic behaviors
\begin{eqnarray}
F_{\rm Ai}(z) \approx
\begin{cases}
&\frac{1}{\pi}\, \sqrt{|z|}\quad\quad\quad {\rm as}\quad 
z\to 
-\infty \label{F1_asymp_left} \\
& \frac{1}{8\pi z}\, e^{-\frac{4}{3}\, z^{3/2}} \;\;\; {\rm 
as}\quad z\to +\infty\, .
\label{F1_asymp_right}
\end{cases}
\end{eqnarray}
Far to the left of the right edge, using
$F_{\rm Ai}(z)\sim \sqrt{|z|}/\pi$ as $z \to -\infty$ in Eq. 
(\ref{F1_asymp_left}), it is easy to show that  
the scaling form (\ref{edge_density_1d}) smoothly matches with 
the semi-circular density in the bulk (\ref{wigner}).   

Another important application of this scaling form (\ref{airy_kernel.1}), combined with the formula in Eq. (\ref{fred_xmax}), is the 
expression of the cumulative distribution of the position of the rightmost fermion $x_{\max}(T=0)$ among $N$ noninteracting fermions
in a harmonic trap at $T=0$. Using the expression (\ref{fred_xmax}) specified with $M=x_{\rm edge}+s \,w_N$, one obtains that the typical quantum fluctuations of $x_{\max}(T=0)$, correctly centered and scaled, 
are governed by the celebrated Tracy-Widom (TW) distribution for GUE, ${\cal F}_2(x)$ \cite{TW94}. Indeed one has
\begin{eqnarray} \label{xmax_0T}
x_{\max}(T=0) = x_{\rm edge} + w_N \, \chi_2 \;,
\end{eqnarray}
where the cumulative distribution function (CDF) of the random variable $\chi_2$ is ${\cal F}_2(s) = {\Pr}(\chi_2 \leq s)$, which can be written as
a Fredholm determinant (\ref{fred_xmax})
\begin{eqnarray}\label{fredholm_F2} 
{\cal F}_2(s) = {\rm Det}(I - P_s {\cal K}_{\rm Ai} P_s) \;,
\end{eqnarray} 
where ${\cal K}_{\rm Ai}(z,z')$ is the Airy kernel given in Eq. (\ref{airy_kernel.1}) and $P_s$ is a projector on the interval $[s,+\infty)$. Note that ${\cal F}_2(s)$ can also be written in terms of a special solution $q(x)$ of the following Painlev\'e II equation \cite{TW94}
\begin{eqnarray}\label{PII}
q''(x) = x q(x) + 2 q^3(x) \;, \; q(x) \sim {\rm Ai}(x) \;, \; x \to \infty \;. 
\end{eqnarray}
The TW distribution ${\cal F}_2(s)$ can then be expressed as
\begin{eqnarray}
{\cal F}_2(s) = \exp{\left[- \int_s^\infty (x-s) q^2(x) \, dx \right]} \;.
\end{eqnarray}
In particular its asymptotic behaviors are given by \cite{BBD08}
\begin{eqnarray} \label{asympt_TW}
{\cal F}_2(s) \sim
\begin{cases}
&\tau_2 \dfrac{e^{-\frac{1}{12}|s|^3}}{|s|^{1/8}}\left(1+ \dfrac{3}{2^6 |s|^3} + {\cal O}(|s|^{-6})\right) \;, \; s \to - \infty \;, \\
& \\
& 1 - \dfrac{e^{-\frac{4}{3} s^{3/2}}}{16\pi s^{3/2}} \left(1-\dfrac{35}{24 s^{3/2}} + {\cal O}(s^{-3}) \right) \;, \; s \to + \infty \;,
\end{cases}
\end{eqnarray}
where $\tau_2 = 2^{1/24}e^{\zeta'(-1)}$ where $\zeta'(x)$ is the derivative of the Riemann zeta function. Quite remarkably, the TW distribution appears in a wide variety of systems, however this free fermion problem is certainly one of the simplest where it naturally arises.

These two examples, (\ref{edge_density_1d}) and (\ref{fredholm_F2}), illustrate how the results and tools from RMT
can be transposed to study the edge properties of the Fermi gas, which are otherwise very hard to study using standard methods like LDA. This has been recently exploited to study other physical properties
like the number variance, i.e. the variance of the number of fermions inside a box \cite{marino_prl,marino_pre}, order statistics \cite{perez-castillo} as well as the entanglement entropy of a trapped Fermi gas \cite{CLM15}.

Here we have discussed the case of a pure harmonic potential. However, one can show \cite{fermions_review,Eis2013,Bornemann2016} that the scaling forms found both in the bulk (\ref{k_bulk}) as well as at the edges (\ref{k_soft_edge}, \ref{airy_kernel.1}) actually hold for a wide class of smooth confining potentials of the form $V(x) \sim |x|^p$ with $p>0$. The $N$-dependence of the length scales $\ell(x)$ (\ref{k_bulk}) and $w_N$ (\ref{wN}) will depend explicitly on $p$ but the scaling functions, namely the sine kernel (\ref{sine_k}) and the Airy kernel (\ref{airy_kernel.1}), are universal. From the point of view of RMT, the universality at the edge is somewhat expected. Indeed, for such potentials $V(x) \sim |x|^p$, Eq. (\ref{rho_LDA}) predicts that the density does have a square-root singularity at the edge, and therefore one would therefore expect that the correlations at the edge are governed by the Airy kernel \cite{Kuij}.

\subsection{Hard box potential (JUE) and the hard-edge kernel}

For the hard-box potential given in (\ref{hb_pot_1d}), setting $R=1$, the bulk density (\ref{rho_LDA}) is uniform inside the box and it is simply given by
\bea \label{unif_density}
\rho_N(x) \approx  \frac{2m}{N \pi \hbar} \theta(1-|x|) = \frac{k_F}{N \pi} \theta(1-|x|) \;,
\eea
where we have introduced $k_F = \sqrt{2m\mu}/\hbar$ -- we recall that $\mu$ is the Fermi energy, i.e., in this case $\mu = \epsilon_N = \hbar^2 \pi^2 N^2/(8m)$ [see Eq. (\ref{hb_1d_wf})]. In fact, in this problem there is a single length scale~(\ref{k_bulk})
\bea \label{ell_unif}
\ell \equiv \ell(x) = \frac{1}{k_F} \:,
\eea
which characterizes the fluctuations both in the bulk and at the edge. Here the density exhibits two edges at $x = \pm x_{\rm edge}$ with $x_{\rm edge} = 1$ but, in contrast to the harmonic potential (\ref{wigner}), here the edges are {\it hard}. In particular, since the eigenfunctions vanish at $x = \pm 1$, i.e. $\phi_l(x=\pm 1) = 0$, it follows from Eq. (\ref{coinciding}) that the density is strictly zero at the edges, i.e., $\rho_N(x = \pm 1) = 0$ for all $N \geq 1$. This also indicates that the formula above (\ref{ell_unif}) will fail to hold sufficiently close to the boundaries $x = \pm 1$ of the box. In fact, close to the hard edges at $x = \pm 1$, the kernel takes the scaling form (say near the right edge $x = + 1$)
\bea \label{hb_scaling}
K_\mu(x,y) \approx k_F {\cal K}_{\rm Hb}\left(k_F(1-x), k_F(1-y) \right) \;,
\eea  
where ${\cal K}_{\rm Hb}(z,z')$ is a hard-box kernel \cite{LLMS17, LLMS18,CMV2011}
\bea\label{hb_form}
{\cal K}_{\rm Hb}(z,z') = \frac{\sin{(z-z')}}{\pi(z-z')} - \frac{\sin(z+z')}{\pi(z+z')} \;,
\eea
which is actually a special case of the so-called Bessel kernel, well known in RMT \cite{forrester} (see also below). Note that the structure of the hard-box kernel (\ref{hb_form}), ${\cal K}_{\rm Hb}(z,z') = {\cal K}_{\rm Sine}(z-z') - {\cal K}_{\rm Sine}(z+z')$, where ${\cal K}_{\rm Sine}(z)$ is the sine-kernel (\ref{sine_k}) indicates that this kernel can be actually obtained by the method of images \cite{LLMS17, LLMS18}. From these results (\ref{hb_scaling}) and (\ref{hb_form}), together with (\ref{coinciding}) we obtain the density profile near the edge at $x=+1$ (a similar formula holds at the left edge)
\bea \label{density_hb}
\rho_N(x) \approx \frac{k_F}{N} F_{\rm Hb}(k_F(1-x)) \;, 
\eea  
where the scaling function $F_{\rm Hb}(z)$ reads
\bea \label{F_Hb_explicit}
F_{\rm Hb}(z) = \frac{1}{\pi}\left(1 - \frac{\sin(2z)}{2z}\right) \;.
\eea
Its asymptotic behaviours are given by 
\bea \label{F_Hb}
F_{\rm Hb}(z) = 
\begin{cases}
&\dfrac{2}{3 \pi}z^2 + {\cal O}(z^4) \\
& \\
& \dfrac{1}{\pi} + {\cal O}(z^{-1}) \;.
\end{cases}
\eea
It describes the crossover from the vanishing density at the boundary (corresponding to the limit $z \to 0$ in (\ref{F_Hb})) to the  
constant density profile in the bulk (described by the $z \to \infty$ limit in (\ref{F_Hb})). The hard-box kernel (\ref{hb_form}) corresponds here to the hard-edge scaling limit of the Jacobi Unitary ensemble (\ref{JUE_gen}) with the special value of the parameters $a=b=1/2$. Different values of the parameters $a,b$, associated to the quantum potential in Eq.~(\ref{pot_JUE}) yield different Bessel kernels with an index that depends on $a$ (at the left edge $\theta = 0$) and $b$ (at the right edge $\theta = \pi$)~\cite{LLMS18}. We refer the reader to Ref. \cite{Cunden1D} for the study of different boundary conditions imposed on the wave-functions $\phi_l(x)$ at the boundary of the box (for instance periodic or Neumann), for which the corresponding joint PDF relates to different unitary matrix models.

\subsection{The potential $V(x) = A\,x^2 + B/x^2$ (LUE) and the Bessel kernel}\label{Sec:LUE}

We now consider the case where $V(x)$ is given in Eq. (\ref{def_inv_sq}), which corresponds to 
the LUE (\ref{joint_LUEsq}) -- we recall that here we set $m = \hbar = 1$. For large $N$, the Fermi energy
$\mu = \epsilon_N$ [see Eq. (\ref{ev_laguerre})] behaves as $\mu \approx 2 b N$. Hence, from Eq. (\ref{rho_LDA}), 
the bulk density is given by 
\be\label{density_inv_sq}
\rho_N(x) \approx \sqrt{\frac{b}{2N}} \tilde f_W\left(x \sqrt{\frac{b}{2N}}\right) \;, \;\; \tilde f_W(z) = 2 \theta(z) f_W(z) = \theta(z) \frac{2}{\pi}\sqrt{2-z^2} \;,
\ee
where $\tilde f_W(z)$ is thus the ``half'' Wigner semi-circle. Note that one can easily check from (\ref{density_inv_sq}) that the variable $y = x^2$ is distributed
according to the Mar\v{c}enko-Pastur law, as expected from the mapping to the LUE in Eq. (\ref{joint_LUEsq}). The bulk density (\ref{density_inv_sq}) exhibits two edges: a ``soft'' edge at $x = 2 \sqrt{N/b}$ and a ``hard'' edge at $x=0$. Near the soft edge, the limiting kernel, properly scaled, is given by the Airy kernel studied above (\ref{airy_kernel.1}). However, near the hard edge at $x=0$, the kernel is different and given by the so-called Bessel kernel, also well known in RMT \cite{forrester,TWBessel}. Indeed, one has (see for instance \cite{LLMS18} in the context of fermions) 
\bea \label{bessel_kernel}
K_\mu(x,y) \approx 2 k_F^2 \sqrt{x\,y} \, {\cal K}_{\rm Be, \alpha-1/2}(k_F^2 x^2, k_F^2 y^2) \;,
\eea
where $k_F = \sqrt{2 \mu}$ and ${\cal K}_{\rm Be,\nu}(u,v)$ is the Bessel kernel of index $\nu$, given by
\bea \label{bessel_explicit}
{\cal K}_{\rm Be,\nu}(u,v) = \frac{\sqrt{v} \, {\rm J}'_{\nu}(\sqrt{v}) {\rm J}_\nu(\sqrt{u}) - \sqrt{u} \, {\rm J}'_\nu(\sqrt{u}) {\rm J}_\nu(\sqrt{v})}{2(u-v)} \;,
\eea
where ${\rm J}_\nu(x)$ denotes the Bessel function of index $\nu$. Note that in the limit $\alpha \to 1$, one can check that 
Eqs. (\ref{bessel_kernel}) and (\ref{bessel_explicit}), using ${\rm J}_{1/2}(z) = \sqrt{2/(\pi z)}\sin{z}$, yield the hard-box kernel (\ref{hb_form}), as expected by comparing the Hamiltonians in Eq. (\ref{hb_pot_1d}) and (\ref{def_inv_sq}) -- since close to the wall at $x=0$ the quadratic term in (\ref{def_inv_sq}) plays no role, at leading order for large $N$. From Eq. (\ref{coinciding}), by evaluating the kernel (\ref{bessel_explicit}) at coinciding points, we obtain the density profile near the hard edge at $x=0$, 
\bea \label{density_bessel}
\rho_N(x) \approx \frac{k_F}{N} F_{\rm Be}(k_F x) \;, \; 
\eea 
with the scaling function
\bea \label{explicit_profile}
F_{\rm Be}(z) =  \frac{z}{2} \left( {\rm J}^2_{\alpha-1/2}(z) - {\rm J}_{\alpha+1/2}(z) {\rm J}_{\alpha-3/2}(z)\right) \;.
\eea
It behaves asymptotically as
\bea\label{bessel_asympt}
\hspace*{-0cm}F_{\rm Be}(z) \sim
\begin{cases} 
& A_\alpha \, z^{2\alpha} \;, \; z \to 0\\
& \dfrac{1}{\pi} \;, \;\;\;\;\;\;\;\;\;\; z \to \infty \;,
\end{cases}
\eea
where $A_\alpha = [2^{2\alpha}\Gamma(\alpha+3/2) \Gamma(\alpha+1/2)]^{-1}$, with $\Gamma(z)$ denoting the Gamma-function. The scaling function $F_{\rm Be}(z)$ interpolates between a vanishing density exactly at the wall and a constant value $\rho_N(x) \approx (2/\pi) \sqrt{b/N}$ far from the wall, i.e. for $k_F x \gg 1$ (but still $x\ll 1$), which matches perfectly with the bulk density in Eq.~(\ref{density_inv_sq}) for $x \to 0$.

\subsection{{General power law potentials $V(x) \sim 1/|x|^\gamma$ : transition from hard edge to soft edge}}

Since it is now possible to realize virtually any confining potential in cold atom experiments
\cite{BDZ08,Zwi2017,2DBoxGaz}, it is an
interesting question to ask about the edge universality for potentials which diverge at the
boundary with some arbitrary power law. We will restrict the discussion to one dimension. 
We already know that hard box potentials relate to the JUE and $1/x^2$ to the LUE, so
we ask now about a potential of the form $V(x) \sim 1/|x|^\gamma$ with $\gamma>0$.

First one must ask whether a barrier $V(x) \sim 1/|x|^\gamma$ is actually confining. It turns out
that at the level of a single particle, if $0<\gamma<1$ the barrier is penetrable, i.e. there exists eigenstates which do not vanish at $x=0$. If $\gamma \leq 1$ the barrier is impenetrable, i.e. the eigenfunctions vanish at $x=0$ \cite{Andrews}. 
In the first case the potential can not act as a trap for
a system of $N$ noninteracting fermions, hence we restrict to the second case, $\gamma>1$. In this case with no loss of generality, one can restrict to $x>0$ and assume that $V(x)=+\infty$ for
$x<0$. 

In Ref. \cite{LLMS18} it was shown that for $1 \leq \gamma < 2$ the barrier acts {\it as an infinite hard wall}. 
In particular, in the solvable case $\gamma=1$ one has \cite{LLMS18} 
\bea
K_\mu(x,y) = k_F K(k_F x ,k_F y;k_F) 
\eea
where $\mu=\q k_F^2$, in terms of a non-trivial reduced kernel $K$, which in
the large $\mu$ limit converges to 
\be
K(z, z';\infty) = \frac{\sin (z-z')}{\pi(z-z')} - \frac{\sin (z+z)}{\pi(z+z')} =  K_{\rm Hb}(z,z') \;,
\ee 
which is the hard box kernel given in \eqref{hb_form}. One can further show \cite{LLMS18} 
that this limiting kernel actually holds for all values $1 \leq \gamma < 2$. This can be understood qualitatively as
follows: the position of the edge density is $r_{\rm e}= k_F^{-2/\gamma}$. Hence for $\gamma <2$ the scaled position of the edge, $k_F r_{\rm e}$
tends to zero at large $\mu$. This is why the edge universality 
is the one of the hard wall for $1 \leq \gamma <2$. More detailed arguments can be found in~\cite{LLMS18}.

For the special value $\gamma=2$ we already know the answer from Section \ref{Sec:LUE} and the edge universality is in this case
the one of LUE, i.e. the Bessel kernel $K_{\rm Be,\nu}$ with an index 
$\nu$ that depends continuously on the {\it amplitude} of the potential [see Eq. (\ref{bessel_kernel})]. It is thus a marginal case. The more surprising fact is that for $\gamma>2$, i.e. very steep potentials, {\it one recovers the Airy universality class}, i.e. the soft edge RMT results ! One way to see it is to consider the case $\gamma=2$, i.e. $V(x) = \alpha(\alpha-1)/(2x^2)$ as in Eq.~(\ref{def_inv_sq}) with $b=0$ (since the quadratic term is irrelevant close to the origin) in the limit of large amplitude $\alpha \to +\infty$. The limiting kernel near the hard edge, given by the Bessel kernel in Eq. (\ref{bessel_kernel}), can be analysed in the large $\alpha$ limit by using the asymptotic result 
\be \label{convergence} 
\lim_{\nu \to +\infty} 2^{2/3} \nu^{4/3} K_{\rm Be,\nu}(\nu^2 + 2^{2/3} \nu^{4/3} \tilde a,
\nu^2 + 2^{2/3} \nu^{4/3} \tilde b) = {\cal K}_{\rm Ai}(- \tilde a , - \tilde b) \;,
\ee 
where ${\cal K}_{\rm Ai}(z,z')$ is the Airy kernel (\ref{airy_kernel.1}). In this case, one has, for large $\alpha$, $k_F r_{\rm e} \simeq \alpha$ \cite{LLMS18}. Hence we will center
the kernel around this point and define
\be
x= r_{\rm e} + w_N \tilde x \quad , \quad y= r_{\rm e} + w_N \tilde y
\ee 
with $k_F w_N= 2^{-1/3} \alpha^{1/3}$. The convergence
property \eqref{convergence} then leads \cite{LLMS18} to the following behaviour of the kernel close to the edge $|x-r_{\rm e}|/w_N\sim |y-r_{\rm e}|/w_N=O(1)$
 \be\label{airy_k_largea}
 K_{\mu}(x,y) \approx \frac{1}{w_N}{\cal K}_{\rm Ai}\left(\frac{r_{\rm e}-x}{w_N},\frac{r_{\rm e}-y}{w_N}\right)\;\;{\rm with}\;\;{\cal K}_{\rm Ai}(z,z')=\int_0^{\infty}du \Ai(z+u)\Ai(z'+u)\;.
 \ee
For fixed $\gamma>2$ one can indeed argue that the Airy universality class holds, see discussion in Appendix A of 
\cite{fermions_review} (below Eq. (A28)). 

In conclusion, to realize the hard wall universality class at the edge in experiments one should use potentials
with $1 \leq \gamma <2$. Contrarily to naive expectations, a more strongly divergent potential
does not lead to the hard wall. Instead, by increasing $\gamma$, one goes from hard wall (JUE), 
to Bessel (LUE), and then finally to Airy (GUE) classes.

\section{Noninteracting trapped fermions in $d$-dimensions at $T=0$}\label{ldim}

Until now, we have focused on noninteracting trapped fermions in one dimension, $d=1$, and we have shown that
there are strong connections between these models and classical unitary matrix models of RMT (GUE, JUE and LUE). 
But it is natural to ask what happens in higher dimensions $d \geq 2$ and consider $N$ spin-less noninteracting fermions in a 
$d$-dimensional potential $V({\bf x})$, with ${\bf x} \in {\mathbb R}^d$. The model is then described by an $N$-body Hamiltonian $\hat 
{\cal H}_N = \sum_{j=1}^N \hat h_j$, where $\hat h_j= \hat h({\bf p}_j,{\bf x}_j)$ is a 
one-body Hamiltonian of the form $\hat h \equiv \hat h({\bf p}_j,{\bf x}_j)$ with
\bea \label{H} 
\hat h = \frac{\hat {\bf p}^2}{2m}  + V({\bf x})  \quad , 
\quad \hat {\bf p} = \frac{\hbar}{i} \mathbf{\nabla} \;. \label{defhhat}
\eea 
Here, for simplicity, we will consider the isotropic $d$-dimensional 
harmonic oscillator
\bea \label{HO} 
V({\bf x}) \equiv V(r) = \frac{1}{2} m \omega^2 r^2 \quad , \quad r = |{\bf x}| \;,
\eea 
which can be solved exactly, but other types of potential, e.g., spherical hard-box potentials can also be studied exactly (see Refs. \cite{LLMS17,LLMS18} for more details). In higher dimensions $d\geq 2$, there is no direct mapping to random matrix models (except in some very special cases in $d=2$, see e.g. \cite{LMS18}). In spite of this, the positions of the fermions form a $d$-dimensional determinantal point process, for which there exist powerful analytical tools, which allow in particular to study the edge properties of the Fermi gas \cite{fermions_review}. Here we will briefly recall the main results obtained along this line, and we refer the reader to \cite{fermions_review} for more details about this higher-dimensional case.

\subsection{Zero temperature statistics}
At zero temperature, as in the $1d$ case (\ref{Slater_1d}), the ground state many-body wave function $\Psi_0$ can be expressed as 
an $N \times N$ Slater determinant, 
\bea \label{slater} 
\Psi_0({\bf x}_1, \cdots, {\bf x}_N) = \frac{1}{\sqrt{N!}} \, 
\det[ \psi_{{\bf k}_i} ({\bf x}_j)]_{1\leq i,j \leq N} 
\eea 
constructed from the $N$ single 
particle wave functions labeled by a sequence $\{ {\bf k}_i \}$, $i=1, \ldots, N$, with 
non-decreasing energies such that $\epsilon_{{\bf k}_i} \leq \mu$ where $\mu$ is the Fermi 
energy. For the isotropic harmonic oscillator, the energy levels and corresponding eigenfunctions are given 
\bea \label{spectrumHO}  
\epsilon_{\bf k}=\sum_{a=1}^d \left(k_a + \frac{1}{2}\right) \hbar 
\omega \quad , \quad \psi_{\bf k}({\bf x})= \prod_{a=1}^d 
\phi_{k_a}(x_a) \;,
\eea 
where the $k_a$'s are integers which range from $0$ to $\infty$, and where $\phi_k(x)$ are the
single-particle eigenfunctions for the one-dimensional harmonic oscillator (\ref{def_Hermite}). 
Note that, in general, he $N$-body ground state is degenerate (whenever the last single 
particle level is not fully occupied). However, since the effect of degeneracy is subdominant
in the large $N$ limit that we are interested in \cite{fermions_review}, we will assume here
that the last level is fully occupied. In this case, for the 
harmonic oscillator, by filling up completely the levels up to $\mu$, one obtains $N = 
\sum_{{\bf k} \in \mathbf{Z}^d} \theta\left(\mu - \hbar \omega(k_1+..+ k_d)\right)$, where we recall that $\theta(x)$ 
is the Heaviside theta function. This leads for large $N$, to 
\begin{equation}
\mu \simeq \hbar \omega 
[\Gamma(d+1) \, N]^{1/d}.\label{musho}
\end{equation}
Using exactly the same manipulations as in $1d$ [see Eqs. (\ref{jpdf_2}, \ref{jpdf_3}, \ref{def_kernel_1d})], the quantum probability, given by the 
squared many-body wave function, can be written as
\begin{eqnarray}\label{eq:psi_0_d}
\hspace*{-2cm}P_{\rm joint}({\bf x}_1, \cdots, {\bf x}_N) =  |\Psi_0({\bf x}_1, \cdots, {\bf x}_N)|^2&=& \frac1{N!} \, \det[ \psi^*_{{\bf k}_i}({\bf x}_j)]  \det[ \psi_{{\bf k}_i}({\bf x}_j)] \\
&=& \frac{1}{N!} \det_{1\leq i,j \leq N} K_\mu({\bf x}_i,{\bf x}_j)
\end{eqnarray} 
in terms of the $d$-dimensional kernel
\begin{equation}\label{eq:def_kernel_d} 
K_\mu({\bf x},{\bf y}) =\sum_{\bf k} \theta(\mu-\epsilon_{\bf k}) 
\psi_{\bf k}^*({\bf x})\psi_{\bf k}( {\bf y}) \;.
\end{equation}
One can easily check that the kernel satisfies the reproducing property, i.e. the $d$-dimensional generalisation of (\ref{self_reprod_1d}), which
follows straightforwardly from the orthonormality of the single-particle eigenfunction. From this reproducing property, together with the determinantal structure of the
quantum joint PDF in (\ref{eq:psi_0_d}), it follows that the $n$-point correlation functions $R_n({\bf x}_1,\cdots, {\bf x}_n)$  can be written as determinants, 
\begin{eqnarray}\label{eq:def_Rn_d}
R_n({\bf x}_1,\cdots, {\bf x}_n) &=& 
\frac{N!}{(N-n)!} \int d{\bf x}_{n+1} \cdots d {\bf x}_N \, 
P_{\rm joint}({\bf x}_1, \cdots, {\bf x}_N) \\
&=&  \det_{1\leq i,j \leq n} K_\mu({\bf x}_i,{\bf x}_j) \;.
\end{eqnarray}
And in particular, for $n=1$, this yields the average density of fermions (normalised to unity)
\bea\label{density_d}
\rho_N({\bf x})= \frac{1}{N} \sum_{i=1}^N \langle \delta({\bf x}-{\bf x}_i) \rangle_0  = \frac{1}{N}R_1({\bf x}) = \frac{1}{N}\, K_\mu({\bf x},{\bf x}) \;,
\eea
where, here, $\langle \cdots \rangle_0$ stands for an average over the joint PDF in Eq. (\ref{eq:psi_0_d}). This result for the correlations (\ref{eq:def_Rn_d}) explicitly shows that the positions of the noninteracting trapped fermions form a $d$-dimensional determinantal point process. This, in turn, implies that all the information
about the correlations is thus contained in the kernel $K_\mu({\bf x}, {\bf y})$, which we now study in the limit of large $N$.  
\subsection{Bulk properties}

For $N \gg 1$, the bulk density can be computed from the LDA approximation, i.e. the straightforward generalisation of (\ref{rho_LDA}) to $d$-dimensions. As in the $1d$-case (\ref{wigner}), it has a finite support 
\be
\rho_N({\bf x}) \approx {1\over 2^d N \pi^{d\over 2} \Gamma(1+{d\over 2})}\alpha^{2d}\left(r^2_{\rm edge} -r^2\right)^{d\over 2} \theta \left(r_{\rm edge} -r\right) \;, \; r = |{\bf x}|\;,\label{rhosho}
\ee
with an extended edge (namely a $(d-1)$-sphere, see Fig. \ref{Fig_edge}) located at 
\be \label{r_edge}
r_{\rm edge} = \frac{2^{1\over 2} [\Gamma(d+1)]^{1\over 2d}}{\alpha}N^{1\over 2d} \;,
\ee
which generalises the Wigner semi-circle found in $d=1$ (\ref{wigner}). Therefore, the kernel behaves differently in the bulk and at the edge. If we consider two points ${\bf x}'$ and ${\bf y}'$ both close to a point ${\bf x}$ in the bulk, the kernel takes the scaling form
\begin{eqnarray}\label{scaling_bulk}
K_{\mu}({\bf x}+ {\bf x}',{\bf x}+{\bf y}') \approx \frac{1}{\ell({\bf x})^{d}} {\cal K}^{\rm bulk}_d\left(\frac{|{\bf x'}-{\bf y'}|}{\ell({\bf x)}}\right)
\end{eqnarray}
where 
\be \label{lx} 
\ell({\bf x}) = \frac{1}{2}\,[N \rho_N({\bf x}) \gamma_d]^{-1/d}  \quad , \quad {\rm with} \;\; \gamma_d =\pi^{d/2} [\Gamma(d/2+1)]
\ee
is the typical local separation between fermions in the bulk. The explicit formula for the scaling function in Eq. (\ref{scaling_bulk}) is given by \cite{Castin, fermions_review,torquato}
\begin{eqnarray} \label{eq:kernel_bulk2}
{\cal K}^{\rm bulk}_d(z) = \frac{{\rm J}_{d/2}(z)}{(2\pi z)^{d/2}} \;,\label{bksho}
\end{eqnarray}
where  ${\rm J}_\nu(z)$ is the Bessel function of index $\nu$. Note that this scaling function (\ref{bksho}) 
has a well defined limit at the origin with ${\cal K}^{\rm bulk}_d(0)= 1/(2^{d}\gamma_d)$. In $d=1$, using
${\rm J}_{1/2}(z) = \sqrt{2/(\pi z)} \sin{z}$, we recover the standard sine-kernel ${\cal K}_{d=1}^{\rm bulk}(z) = {\cal K}_{\rm Sine}(z)= {\sin z}/{(\pi z)}$ of RMT (\ref{sine_k}). While the standard derivation of this result for the bulk kernel (\ref{eq:kernel_bulk2}) usually relies on the LDA \cite{Castin}, we provided a more controlled derivation of it using a method relying on the representation of the kernel in terms of the quantum propagator (in imaginary time) \cite{fermions_review}. As in $1d$, this form of the kernel (\ref{scaling_bulk}, \ref{lx}, \ref{eq:kernel_bulk2}) holds in the bulk, where the density is finite, but breaks down at the edge, for $|{\bf x}| = r_{\rm edge}$, where the density vanishes. 
\subsection{Edge statistics}
Near the edge, the limiting kernel is described by a different scaling form \cite{DPMS:2015,fermions_review}
\begin{eqnarray}\label{eq:def_edge_kernel}
K_{\mu}({\bf x},{\bf y}) \approx \frac{1}{w_N^d} {\cal K}^{\rm edge}_d\left(\frac{{\bf x} - {\bf r}_{\rm edge}}{w_N},\frac{{\bf y} - {\bf r}_{\rm edge}}{w_N}\right) \;.
\end{eqnarray}
To give an explicit expression of the scaling function at the edge, it is useful to split any vector ${\bf v}$ as ${\bf v} = {\bf v}_t + v_n {\bf n}$ where ${\bf n}={\bf r}/r$ is the direction normal to the edge. With these notations, the scaling function reads \cite{DPMS:2015,fermions_review}
\bea\label{eq:edge_kernel3}
{\cal K}_d^{\rm edge}({\bf a},{\bf b})=\int\frac{d^{d-1}{\bf l}}{(2\pi)^{d-1}}\e^{i{\bf l}\cdot({\bf a}_t-{\bf b}_t)}\int_{{\bf l}^2}^{\infty}dz \Ai(a_n+z)\Ai(b_n+z)\;.
\eea
In fact, the integral over the angular variables can be performed explicitly \cite{LLMS18} and this yields finally (for $d \geq 2$)
\be\label{eq:edge_kernel4}
{\cal K}_d^{\rm edge}({\bf a},{\bf b}) = \int_0^{\infty} dl \left(\frac{l}{2\pi}\right)^{\frac{d-1}{2}}\frac{\J_{\frac{d-3}{2}}(l|{\bf a}_t-{\bf b}_t|)}{|{\bf a}_t-{\bf b}_t|^{\frac{d-3}{2}}}\int_{l^2}^{\infty}dz \Ai(a_n+z)\Ai(b_n+z)\;.
\ee
This kernel is a generalisation of the Airy-kernel (\ref{airy_kernel.1}) and, in particular, in Eq. (\ref{eq:edge_kernel3}), it can be seen that ${\cal K}_{d = 1}^{\rm edge}({a},{b}) = {\cal K}_{\rm Ai}(a,b)$, in agreement with the $1d$ result (\ref{airy_kernel.1}). Note that this limiting kernel (\ref{eq:edge_kernel3}) was obtained using the method of the quantum propagator in imaginary time, and not directly from the formulae given in Eqs. (\ref{spectrumHO}) and (\ref{eq:def_kernel_d}) -- see \cite{fermions_review} for more details. 

Here we have discussed the $d$-dimensional (purely) harmonic oscillator $V({\bf x}) = m \omega^2 |{\bf x}|^2/2$ (\ref{HO}) but one can show that the kernels both in the bulk (\ref{eq:kernel_bulk2}) and at the edge (\ref{eq:edge_kernel4}) are actually universal and hold for smooth spherically symmetric potentials with a single minimum, such as $V({\bf x}) \sim |{\bf x}|^p$ for $p>0$ (see \cite{fermions_review} for a more precise statement about universality in this case). In this case, the typical width of the edge regime scales like $w_N \sim N^{-2(p-1)/(3d(p+2))}$.

\subsection{Extremal statistics}
As in the one-dimensional case (\ref{xmax_0T}), it is also interesting to investigate extreme value questions for such $d$-dimensional determinantal point processes (\ref{eq:def_Rn_d}). In particular, in $1d$, the distribution of the position of the rightmost particle $x_{\max} = \max\{x_1, x_2, \cdots, x_N \}$, properly shifted and scaled, converges for large $N$ to the celebrated Tracy-Widom distribution (\ref{xmax_0T}, \ref{fredholm_F2}). For $d$-dimensional system, a natural extreme value 
observable is the maximal radial distance of the fermions from the trap center, defined as \cite{farthest_f}
\bea \label{def_rmax}
r_{\max} = \max \{r_1, r_2, \cdots, r_N \} \;, \; {\rm where} \;\; r_i^2 = {\bf x}_i \cdot {\bf x}_i   \;.
\eea
In Ref. \cite{farthest_f} the cumulative distribution of $r_{\max}$, $P(w,N) = {\rm Prob}(r_{\max} \leq w)$, was computed and studied in the large $N$ limit. It was shown that for large $N$ it takes the scaling form
\bea \label{Gumbel}
P(w,N) \approx G\left(\frac{w-A_N}{B_N}\right) \;\;, \;\; {\rm where} \;\; G(z) = e^{-e^{-z}} \;,
\eea
where $A_N$ and $B_N$ can be computed explicitly for large $N$ \cite{farthest_f} [in particular, to leading order for large $N$, $A_N \simeq r_{\rm edge}$ given in Eq. (\ref{r_edge})]. In Eq. (\ref{Gumbel}), the function $G(z)$ is the well known Gumbel distribution, which is one of the limiting distributions that emerges in the classical theory of extreme value statistics of independent and identically distributed (i.i.d.) random variables \cite{gumbel, galambos}. Hence, although the positions of the Fermions are strongly correlated, as a consequence of the Pauli principle, it turns out that, for spherically symmetric potential,  the {\it radial} components of the displacements are actually independent, yielding eventually the Gumbel distribution for the distribution of $r_{\max}$ (\ref{Gumbel}) in the large $N$ limit \cite{farthest_f}. Note that a similar property holds \cite{kostlan} for the moduli of the eigenvalues of random matrices belonging to the complex Ginibre matrices -- which corresponds to another spherically symmetric determinantal point process in dimension $d=2$ \cite{mehta,forrester}. In this case, the largest modulus, properly shifted and scaled, is also distributed according to a Gumbel law, as in (\ref{Gumbel}) \cite{rider}.

\section{Noninteracting trapped fermions in $d$-dimensions at $T>0$}
\label{temp}
We now proceed to study the effects of temperature on $N$ noninteracting fermions
in an external confining potential. Although the analysis can be done for any 
potential and in any arbitrary dimensions, we will focus for simplicity on the harmonic
potential in one dimension, the generalisation to higher dimensions and other potentials
being rather straightforward \cite{fermions_review}. We first focus   
on the canonical ensemble at temperature $T = 1/\beta$, that corresponds to a fixed number of fermions $N$, 
which is often the situation studied in cold atoms experiments.

Let us start with a qualitative scaling analysis to estimate the different relevant temperature scales in the problem, both in the bulk and
at the edge. As soon as temperature is turned on, $T>0$, there is a new length scale in the problem, namely the de Broglie wavelength, given by 
\bea \label{deBroglie}
\lambda_T=\hbar \sqrt{\frac{2 \pi}{m T}} \;.
\eea
The de Broglie wavelength controls the quantum to classical crossover and thus allows us to estimate when the temperature
is relevant, i.e., when it modifies the $T=0$ results found above. In the bulk, the zero temperature characteristic length scale (\ref{k_bulk}) is $\ell(x) = {\cal O}(1/\sqrt{N})$ (see Fig. \ref{Fig_length}). Therefore, when $\lambda_T \gg \ell(x)$, or equivalently $T \ll N$,
the quantum effects are important while quantum fluctuations are irrelevant if  $\lambda_T \ll \ell(x)$, i.e. $T \gg N$. This suggests that the typical temperature scale in the bulk is $T = {\cal O}(N)$. The situation is different at the edge where the characteristic length scale (\ref{wN}) is $w_N =  {\cal O}(N^{-1/6})$ (see Fig. \ref{Fig_length}) and therefore the typical temperature scale, such that $\lambda_T \sim w_N$, is $T = {\cal O}(N^{1/3})$: this clearly shows that the edge is much more sensitive to thermal fluctuations than the bulk. 
\subsection{Canonical and grand canonical statistics}\label{grand_canonical}

To make a more precise analysis at finite temperature, we need to consider all the excited states (not only the ground-state), i.e., the full Hilbert space
of the $N$ particles. A natural basis of this Hilbert space is formed by
the eigenstates of the $N$ particle Hamiltonian $\hat {\cal H}_N$. For noninteracting
fermions, this basis can be constructed from the 
eigenstates $\phi_k(x)$ in (\ref{def_Hermite}) of the single particle Hamiltonian $\hat h$. 
The associated eigenvalues are $\epsilon_k= \hbar\omega(k+1/2)$
where $k$ is an integer which ranges from $0$ to $\infty$. From these single particle eigenstates, one can
construct all many-body eigenfunctions of $\hat {\cal H}_N$ by 
putting $N$ fermions in $N$ different single particle levels indexed by $k_1<k_2<\ldots<k_N$. The corresponding
eigenfunction is given by the Slater determinant $\propto \det_{1\leq i,j \leq N} \phi_{k_i}(x_j)$
built from these single particle levels. In the canonical ensemble, the joint PDF of the particle positions 
is given by the diagonal element of the density matrix $\hat \rho = e^{-\beta \hat {\cal H}_N}/Z_N(\beta)$
where $Z_N(\beta)$ is the canonical partition function. Therefore, in the canonical ensemble, it 
can be written as the Boltzmann weighted sum of such Slater determinants (squared)
\begin{eqnarray}\label{p_start}
\hspace*{-2cm}P_{\rm joint}(x_1, \ldots, x_N) &=& \langle x_1, \cdots, x_N|\hat \rho |x_1, \cdots, x_N\rangle \nonumber  \\
&=&\frac{1}{{N!}Z_N(\beta)}\sum_{k_1< \cdots < k_N} 
\left| \det_{1\leq i,j \leq N} \phi_{k_i}(x_j) \right| ^2 e^{-\beta\, (\epsilon_{k_1}+\cdots+ \epsilon_{k_N})} \;,
\end{eqnarray}
where $Z_N(\beta) = \sum_{k_1<k_2<\ldots< k_N} e^{-\beta\, (\epsilon_{k_1}+\cdots+ \epsilon_{k_N}) }$ is the canonical partition function. 
It is easy to check that $Z_N(\beta)$ is such that the PDF $P_{\rm joint}(x_1, \ldots, x_N)$ 
is normalized to unity. Interestingly, this joint distribution in (\ref{p_start}) turns out to be the joint law of the eigenvalues of a random matrix model, the so-called
Moshe-Neuberger-Shapiro (MNS) model \cite{MNS94}, which has also received some attention in the maths literature \cite{Joh07,Liechty17,JohanssonLambert}. The goal is then to compute $n$-point correlation functions $R_n(x_1, \cdots, x_n)$ defined as in Eq. (\ref{def_correl})
with this joint PDF (\ref{p_start}). However, handling these multiple integrals in (\ref{def_correl}) at finite temperature $T>0$ turns out
to be much more difficult than their counterpart at $T=0$. To appreciate this, we note that the joint PDF in Eq. (\ref{p_start}) can be written as a determinant,
as it is the case for $T=0$,~\cite{MNS94}
\begin{eqnarray}\label{p_start2}
P_{\rm joint}(x_1, \cdots, x_N) &=& \frac{1}{{N!}Z_N(\beta)}
\det_{1\leq i,j \leq N} G(x_i,x_j,\beta \hbar) 
\eea
in terms of the imaginary-time propagator associated to the one-body Hamiltonian 
\bea
G(x,y;t) = \langle y | e^{- \frac{t}{\hbar} \hat h} | x \rangle = \sum_k e^{- \frac{t}{\hbar} \epsilon_k} \phi^*_k(x) \phi_k(y)  \;.
\eea
Unfortunately, and at variance with the $T=0$ case, successive integrations over the coordinates 
$x_i$ do not preserve this determinantal structure. This is because the kernel inside the determinant
no longer satisfies the reproducing property since
\bea
\int_{-\infty}^{\infty} dz \, G(x,z; \beta \hbar)  G(z,y; \beta \hbar) = G(x,y; 2 \beta \hbar)
\eea 
which is clearly a different kernel. Hence the evaluation of these integrals for arbitrary $N$ is very difficult, since the process
is not determinantal in the canonical ensemble at $T>0$.

Fortunately, in the large $N$ limit, it is possible to make further progress by using the equivalence between the thermodynamic ensembles and
work in the grand-canonical ensemble with fixed chemical potential $\tilde \mu$ (which, for $T>0$, is actually different from the Fermi energy $\mu$ discussed previously: $\tilde \mu \to \mu$ only for $T \to 0$). This amounts to considering that the total number of fermions is fluctuating, being itself an exponential random variable with parameter~$\tilde \mu$. The advantage of working in the grand-canonical ensemble is that the positions of the fermions do constitute a determinantal point process.  
Therefore, in the large $N$ limit, the $n$-point correlation functions $R_n(x_1, \cdots, x_n)$ take a determinantal form \cite{fermions_review} (see also \cite{Joh07,hough}) 
\begin{eqnarray}\label{det_process}
R_n(x_1, \cdots, x_n) \approx \det_{1 \leq i,j \leq n} K_{\tilde \mu}(x_i,x_j) \;,
\end{eqnarray} 
where the finite temperature kernel is given by 
\bea \label{kernel_final}
K_{\tilde \mu}(x,x') = \sum_{k=0}^\infty \frac{\phi^*_k(x) \phi_k(x')}{e^{\beta(\epsilon_k - {\tilde \mu})} + 1} \;.
\eea
In Eq. (\ref{kernel_final}) we recognise the Fermi factor $1/(e^{\beta(\epsilon_k - {\tilde \mu})} + 1)$ where the chemical potential ${\tilde \mu}$ is fixed by the relation
\bea \label{rel_mu_N}
N = \sum_{k=0}^\infty \frac{1}{e^{\beta(\epsilon_k - {\tilde \mu})} + 1} \;.
\eea
Let us emphasise that the relation in (\ref{det_process}) is actually exact in the grand-canonical ensemble but only approximate in the canonical ensemble for finite $N$, becoming exact only in the limit $N \to \infty$. As $T \to 0$, using that 
$\tilde \mu \to \mu$ together with the fact that the Fermi factor in Eq. (\ref{kernel_final}) becomes a theta function $\theta(\mu - \epsilon_k)$, we see that the expression in (\ref{kernel_final}) yields back the zero temperature kernel (\ref{def_kernel_1d}). This exact formula for the kernel (\ref{kernel_final}) is then amenable to a large $N$ analysis, which can be carried out both in the bulk and at the edge \cite{us_finiteT,fermions_review}, which, as done for $T = 0$, we discuss separately.

\subsection{Bulk regime at finite temperature} 
Before discussing the two-point kernel, it is useful to analyse the finite temperature density in the bulk, where, as we have seen before [see Eq. (\ref{deBroglie}) and below it], the typical temperature scale is $T = {\cal O}(N)$. For large $N$, one finds that the fermion density $\rho_N(x) = N^{-1} K_{\tilde \mu}(x,x)$ [see Eq. (\ref{coinciding})] in the bulk takes the scaling form \cite{fermions_review}
\bea \label{densityT}
\rho_N(x) \approx \frac{\alpha}{\sqrt{N}} R\left(y = \beta N \hbar \omega, z= x \sqrt{\beta m \omega^2/2} \right) \;,
\eea
with the bulk scaling function
\bea\label{scaling_function}
R(y,z) = - \frac{1}{\sqrt{2 \pi y}} {\rm Li}_{1/2}(- (e^y-1) e^{-z^2}) \;,
\eea
where ${\rm Li}_{1/2}(z) = \sum_{n\geq 1} z^n/n^{1/2}$ is the polylogarithm function of index $1/2$. On this temperature scale, the fermion density  (\ref{scaling_function}) extends over the whole real axis, so that there is no edge for $T = {\cal O}(N)$. Using the asymptotic behaviors, ${\rm Li}_{1/2}(-e^X) \approx -(2/\sqrt{\pi})\, X^{1/2}$, as $X \to \infty$ while ${\rm Li}_{1/2}(X) \approx X$ as $X \to 0$, it is easy to see that $R(y,z)$ interpolates between the Wigner semi-circle (\ref{wigner}) as $T \to 0$ and a Gaussian as $T \to \infty$, namely
\bea \label{Boltzmann}
\rho_N(x) \approx \sqrt{\frac{\beta\,m\, \omega^2}{2\pi}}\, 
\exp\left[-\frac{\beta}{2}\, m\,\omega^2\, x^2\right] \;, \; T \to \infty \;,
\eea
which is just the (classical) Maxwell-Boltzmann weight for independent particles in a harmonic potential $V(x) = (1/2) m\omega^2 x^2$ and at inverse temperature $\beta$. 

The two-point kernel $K_{\tilde \mu}(x,y)$ (\ref{kernel_final}) can also be analysed in the bulk, for $x, y$ both close to the center of the trap, with $x - y = {\cal O}(1/(\alpha \sqrt{N}))$ (the typical inter-particle distance in the bulk) and at finite temperature $T = {\cal O}(N)$. For large $N$, it takes the scaling form 
\bea
K_{\tilde \mu}(x,y) \approx \alpha\, N^{1/2} {\cal K}^{\rm bulk}_{y}\left(\alpha \sqrt{N} (x-y)\right)\; , 
\eea 
where 
\bea\label{Kbulk1d}
{\cal K}^{\rm bulk}_{y}(v) = \frac{1}{\pi \sqrt{2 y}} \int_0^{+\infty} dp \frac{\cos(\sqrt{\frac{2 p}{y}} v)}{(1+ e^p/(e^y-1)) \sqrt{p}} 
\eea 
(see also Refs. \cite{Joh07,Verba} for alternative derivations of this kernel). 

\subsection{Edge regime at finite temperature} 
At the edge, the typical temperature scale is $T = {\cal O}(N^{1/3})$, i.e. $\beta = {\cal O}(N^{-1/3})$ (see Eq. (\ref{deBroglie}) and below). In this scaling regime, the variable $y = \beta N \hbar \omega \gg 1$ for large $N$. Hence the bulk density profile (\ref{scaling_function}) is given by the Wigner semi-circle (\ref{wigner}) in this regime, with two edges at $\pm \sqrt{2N}/\alpha$, as in the case $T=0$. However, the edge kernel is different from the Airy kernel. Indeed, setting
\bea \label{def_b}
b = \frac{\hbar \omega}{T}\, N^{1/3} \;,
\eea 
the kernel $K_{\tilde \mu}(x,y)$ at the edge takes a scaling form similar to the $T=0$ scaling form in Eq. (\ref{k_soft_edge}), i.e.
\bea \label{kff0}
K_{\tilde \mu}(x,y) \simeq \frac{1}{w_N} {\cal K}^{\rm edge}_b \left( \frac{x-x_{\rm edge}}{w_N} , \frac{y-x_{\rm edge}}{w_N} \right) \;,
\eea 
but with a modified scaling function given by \cite{us_finiteT,fermions_review,Joh07}
\begin{eqnarray}\label{kff}
{\cal K}^{\rm edge}_{b}(z,z') = \int_{-\infty}^\infty \frac{{\rm Ai}(z+u){\rm Ai}(z'+u)}{e^{-b\, u} +1} du \;,
\end{eqnarray}
which is a finite temperature generalisation of the Airy kernel ${\cal K}_{\rm Ai}(z,z')$ in Eq. (\ref{airy_kernel.1}). Note that in the limit of zero temperature, when $b \to \infty$, the non-zero contribution to the integral over $u$ on the right hand side of Eq. (\ref{kff}) comes from $u \in [0, +\infty)$ and one gets, using Eq. (\ref{airy_kernel.1}), $\lim_{b \to \infty} {\cal K}^{\rm edge}_{b}(z,z')  = {\cal K}_{\rm Ai}(z,z')$. From this limiting kernel (\ref{kff}) evaluating at $s=s'$ one obtains the finite temperature density profile at the edge, 
\bea
\rho_N(x) \simeq \frac{1}{N w_N} F^{\rm edge}_{b}\left( \frac{x - x_{\rm edge}}{w_N}\right) 
\eea 
where the finite temperature scaling function $F^{\rm edge}_{b}(z)$ is
obtained as
\bea \label{F1bs} 
F^{\rm edge}_{b}(z) = \int_{-\infty}^{+\infty} du \frac{{\rm Ai}(z+u)^2}{1 + e^{-b u}} \;,
\eea 
which thus depends continuously on $b$, through the Fermi factor, and yields back the $T=0$ edge profile (\ref{edge_density_1d}) in the limit $b \to \infty$.  

\begin{figure}[t]
\centering
\includegraphics[width = 0.8\linewidth]{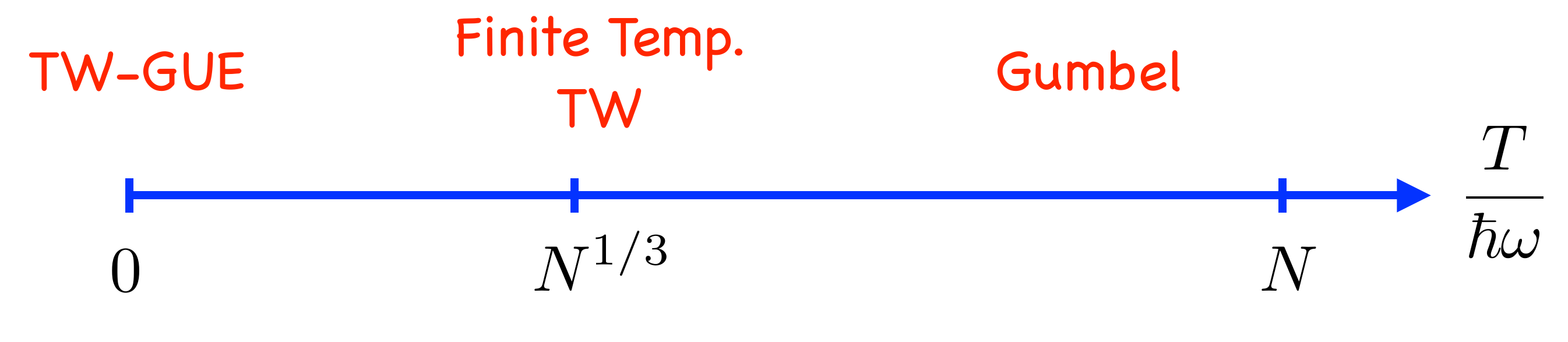}
\caption{Behavior of the distribution of the position of the rightmost fermion $x_{\max}(T)$ as a function of $T$. For $T/(\hbar \omega) \ll N^{1/3}$, the distribution is given by the TW-GUE distribution (\ref{fredholm_F2}), well known in RMT \cite{TW94} while for $T/(\hbar \omega) \gg N^{1/3}$ (and $T/(\hbar \omega) \ll N)$ it behaves as a Gumbel distribution. The full crossover between these two regimes occurs for $T/(\hbar \omega) = {\cal O}(N^{1/3})$ and is described by the ``finite temperature Tracy-Widom'' distribution given in Eq. (\ref{finite_T_Ai}). Note that for $T/(\hbar \omega) \gg N$, the distribution of $x_{\max}(T)$ is described by yet another Gumbel law \cite{fermions_review}, not discussed here.}\label{Fig_FiniteT_TW}
\end{figure}

From the determinantal structure (\ref{det_process}) one can also obtain the distribution of the position of the rightmost fermion at finite temperature, $x_{\max}(T) = \max_{1 \leq i \leq N} x_i$. Indeed, as in the $T=0$ case in (\ref{xmax_0T}) and (\ref{fredholm_F2}), the cumulative distribution of $x_{\max}(T)$, properly shifted and scaled, is given by a Fredholm determinant involving the finite temperature Airy kernel (\ref{kff}). Indeed, one has
\bea \label{finite_T_Ai}
{\Pr \left(x_{\max}(T)  \leq x_{\rm edge} + \frac{N^{-1/6}}{\sqrt{2} \alpha}\,s \right)} = {\rm Det}(I - P_s \,{\cal K}^{\rm edge}_{b}\,P_s) \;,
\eea   
where $P_s$ is the projector on $[s, + \infty)$ and ${\cal K}^{\rm edge}_{b}$ is given in Eq. (\ref{kff}). The Fredholm determinant on the right hand side of Eq. (\ref{finite_T_Ai}) is a finite temperature generalization of the TW distribution, found at $T=0$ (\ref{fredholm_F2}) and it can be expressed in terms of the solution of a non-local Painlev\'e equation \cite{ACQ11} (see also \cite{fermions_review, us_KPZ_largetime, SMP17}). In fact, one can show \cite{Joh07} that this Fredholm determinant interpolates between the TW distribution as $T \to 0$, i.e. $b \to \infty$, and a Gumbel distribution [as in Eq. (\ref{Gumbel})] at high temperature, i.e. $b \to 0$, where the positions of the fermions become completely uncorrelated (see Fig. \ref{Fig_FiniteT_TW}).

 Interestingly, it turns out that the very same Fredholm determinant (\ref{finite_T_Ai}) arises in the exact solution of the Kardar-Parisi-Zhang (KPZ) equation with droplet initial conditions. The origin of this connection remains poorly understood. We refer the interested reader to Refs. \cite{us_finiteT, fermions_review, us_KPZ_largetime,SMP17, us_KPZ_shorttime} for a more detailed discussion as well as further analysis of this Fredholm determinant (\ref{finite_T_Ai}) in the context of large deviations in the KPZ equation, which has recently attracted a lot of attention, both in the physics \cite{us_KPZ_largetime,SMP17, us_KPZ_shorttime, KL2017, Corwin2018,prolhac} and in the maths literature \cite{CT2018,Tsai2018}.


\section{{Correlations in the phase space and edge in momentum space}}\label{Sec:phase_sp}

Here we consider briefly {two} applications and extensions of the methods 
discussed in this review. {Until now we have mostly discussed the 
spatial structure of the correlations of trapped fermions. However, one can also ask about
the correlations in momentum space, $p$, accessible via time of flight experiments 
\cite{2DBoxGaz,TOF}. The tool of choice for exploring the correlations 
phase-space, i.e. in the $(x,p)$-plane, is the Wigner function,
which we first consider. Next, we discuss the possible edge behaviors in momentum space,
which leads to new universality classes.}

\subsection{The  Wigner function of free trapped fermion systems}\label{Sec:Wigner}

In \cite{us_Wigner} the many body Wigner function for trapped systems of non-interacting fermions was studied employing the techniques described in this review. We recall that the Wigner function for a single quantum particle in one dimension is given by \cite{wigner}
\be \label{wigner1}
W_1(x,p) =  \frac{1}{2\pi \hbar} \int_{-\infty}^{+\infty} dy \, e^{i py/\hbar} \psi^*(x + \frac{y}{2}) \psi(x - \frac{y}{2})  \;.
\ee 
where $\psi(x)$ denotes the wave function in the spatial representation. We also denote by $\hat \psi(p) = \int dx\, \psi(x)\,e^{ipx/\hbar}$ the wave function in momentum representation, i.e. the Fourier transform of $\psi(x)$. The Wigner function is a pseudo-probability density function and is often heuristically used as a joint probability distribution function for the position and momentum as its marginal distributions 
are given by
\begin{equation}\label{marginal_Wigner}
\int dp\  W_1(x,p) = |\psi(x)|^2 \ \ \ \ ; \;\;
\int dx\  W_1(x,p) = |\hat \psi(p)|^2,
\end{equation}
that is to say the probability distribution of the position and the probability distribution of the momentum. However, the  Wigner function is referred to as a pseudo-probability density as it is generically non-positive. Despite this, the Wigner function has proved
to be useful in a number of contexts \cite{case}. As mentioned in Section \ref{ds}, it can be used to derive the LDA in bulk systems, but is also used in
quantum chaos and semiclassical physics \cite{berry1,Hannay}, in quantum optics \cite{bookQuantumOptics},
in the theory of optical devices \cite{Bazarov}, in quantum information theory \cite{measurementWigner} or in the context of quantum mirror curves \cite{marcos}. 

For an  $N$ body system in $d$-dimensions and zero temperature, 
the many body Wigner function is defined as 
\bea
W_N({\bf x},{\bf p})  = \frac{N}{(2\pi \hbar)^d} \int_{-\infty}^{+\infty} && d{\bf y} \, d{\bf x}_2 \ldots d{\bf x}_N  \,
e^{\frac{i {\bf p} \cdot {\bf y}}{\hbar}} 
 \Psi_0^*({\bf x}+\frac{{\bf y}}{2}, {\bf x}_2,\ldots, {\bf x}_N)  \nonumber \\
&& \times \Psi_0({\bf x}-\frac{{\bf y}}{2}, {\bf x}_2,\ldots, {\bf x}_N)  \label{wig1def},
\eea
where $\Psi_0({\bf x}_1, {\bf x}_2,\ldots, {\bf x}_N)$ is the ground-state wave function constructed from the Slater determinant. 
This many body Wigner function satisfies
\bea
&& \int_{-\infty}^{+\infty} d\p \, W_N({\bf x},{\bf p})= n_N({\bf x})\ \  ;
  \int_{-\infty}^{+\infty} d{\bf x} \, W_N({\bf x},{\bf p})= \bar \rho_N({\bf p}) \\
&&  \int_{-\infty}^{+\infty} d{\bf x} \, d{\bf p} \, W_N({\bf x},{\bf p})=N \;,
\eea
where $n_N(x) = N\, \rho_N(\x)$ is the average number density of fermions in position space (normalized to $N$), and $\bar \rho_N(\p)$ the average number density in momentum space. 

The classical single particle energy for this system is given by
\be 
E({\bf x},\p) =
 \frac{\p^2}{2 m} + V({\bf x}).  \label{classicalE}
 \ee
The points in phase space $({\bf x}_e,\p_e)$ obeying $E({\bf x}_e,{\bf p}_e)=\mu$, where $\mu$ is the Fermi energy, constitute a semi-classical Fermi surface in classical phase space, known as the {\em Fermi surf} \cite{wiegmann}. This Fermi surf plays the role of the edge for the behaviour of the Wigner function. Within the Fermi surf,  there is a bulk region where one finds
\be
W_N({\bf x},\p)  \simeq \frac{1}{(2 \pi \hbar)^d} \Theta( \mu - E({\bf x},\p) ),  \label{W0bulk}
\ee
that is to say a uniform distribution over the classically permitted phase space. This result can also be obtained via the LDA, see for example \cite{Castin}.  However, this LDA result breaks down near the Fermi surf.

Near a given point on the Fermi surf $({\bf x}_e,{\bf p}_e)$ we can associate an intrinsic energy scale \cite{us_Wigner}
\beq
e_N = \frac{(\hbar)^{2/3}}{(2 m)^{1/3}} \left( \frac{1}{m} (\p_e \cdot  \nabla)^2 V({\bf x}_e)  + |\nabla V({\bf x}_e)|^2 \right)^{1/3} \;, \label{eN_intro} 
\eeq 
which in general will vary over the Fermi surf. Near the point  $({\bf x}_e,{\bf p}_e)$  we can associate the scaled energy variable 
\be
a = \frac{1}{e_N} (E({\bf x},{\bf p}) -  \mu),  \label{ae} 
\ee
which is  a function of only the local classical energy of the point $({\bf x},{\bf p})$. The Wigner function $W_N(\x,\p)$, at $T=0$ and in arbitrary $d$, is then given by
\be
W_N({\bf x},\p)  \simeq \frac{{\cal W}(a)}{(2 \pi \hbar)^d}  \,,\label{scal1}
\ee
where remarkably the scaling function
\be
{\cal W}(a) = \int_{2^{2/3} a}^{+\infty} \Ai(u) du \label{scal2} 
\ee
does not depend on the dimension of space $d$. 

The results at zero temperature can be extended to finite temperature, by again passing over to the grand canonical ensemble (see Section \ref{grand_canonical}). In the bulk we recover the LDA result
\be
W_{\tilde \mu}({\bf x},\p) = \frac{1}{1 + e^{\beta ( \frac{\p^2}{2 m } + V(\bf x) - \tilde \mu)}} \;. \label{WTbulk} 
\ee 
where $\tilde \mu$ is the finite temperature chemical potential which can be obtained from standard statistical mechanics methods \cite{us_Wigner}. Near a given point on the Fermi surf, the Wigner function depends again on its coordinates via the local scaled energy
$a$ defined in Eq. (\ref{ae}) for the zero temperature case and is given by
\be
W_N({\bf x},\p)  \simeq \frac{{\cal W}_b(a)}{(2 \pi \hbar)^d} ,
\ee 
where the scaling function is now given by
\bea
&&   
 {\cal W}_b(a) = \int_{-\infty}^{+\infty} \frac{2^{2/3} du}{1+ e^{-b u}} \Ai(2^{2/3} (u+a)) \; \label{scal2T} 
\eea
and depends on the temperature via the parameter $b$ which is given here by $b = \beta \, e_N$ with $e_N$ given in Eq. (\ref{eN_intro}). This representation is valid in the thermodynamic limit where $e_N(\mu)\to\infty$ while $b$ is of order 1.

\subsection{Edge in momentum space: multicritical universal statistics}

We now ask about the statistics of the {\it momenta} $p_i$ of $N$ noninteracting
fermions, and their maximum $p_{\max}= \max_{i=1, \ldots,N} p_i$, in a (e.g. $1d$) trap described by a single particle Hamiltonian $\hat h=\frac{\hat p^2}{2 m} + V(\hat x)$. From the Wigner 
function bulk formula \eqref{W0bulk},
we see that if the potential is bounded from below
(assume that its minimum occurs at $x=0$ with $V(0)=0$) there exists also an edge in momentum space $p_{\rm e}=\sqrt{2 m \mu}$, beyond which the momentum density
$\bar{\rho}_N(p) = \int dx \, W_N(x,p)$ vanishes. Obviously, if the confining potential is harmonic, i.e., $V(x) = m\omega^2 x^2/2$ 
momenta and positions play a
symmetric role and the two (dimensionless) random sets
$\{p_i/\hbar \alpha \}_{i=1,\ldots,N}$ (momenta) and $\{\alpha x_i\}_{i=1,\ldots,N}$ (coordinates) 
are described by exactly the same joint PDF (here $\alpha = \sqrt{m \omega/\hbar}$
is the harmonic oscillator inverse length scale), at any temperature (and in fact, in any dimension $d$).
This joint PDF is also the one of the GUE eigenvalues, leading to the Airy class at the edge (both in real and
momentum space). This can also be seen from the universality of the
Wigner function, in that case all along the Fermi surf, as discussed in the previous Section.

The question of what happens for a more
general, non harmonic trap, i.e. the pure power law potentials 
$V(x)= g x^{2n}$, with $n \geq 2$ is however, non-trivial. In particular, for $n>1$ the duality between $x$ and $p$ break down.   
From \eqref{W0bulk} we see that the density {\it in momentum space} now vanishes as
\be \label{density} 
\bar{\rho}_N(p) \sim (p_{\rm e}-p)^{\frac{1}{2 n}} \;,
\ee 
i.e, distinct from the standard Wigner 
semi-circle exponent $\frac{1}{2}$ (for $n=1$)
which indicates a new universality class.
The failure of the GUE edge universality can also be seen from the formula
for the width $e_N$ of the edge region [see Eq. \eqref{eN_intro}] associated to the Wigner function.
We see that if the curvature of the
potential vanishes (together with its first derivative)
at the 
point $(x_e=0,p_e)$ in phase space, the formula gives $e_N=0$. This signals
a new universality class.

\begin{figure}
\centering
\includegraphics[width = 0.6\linewidth]{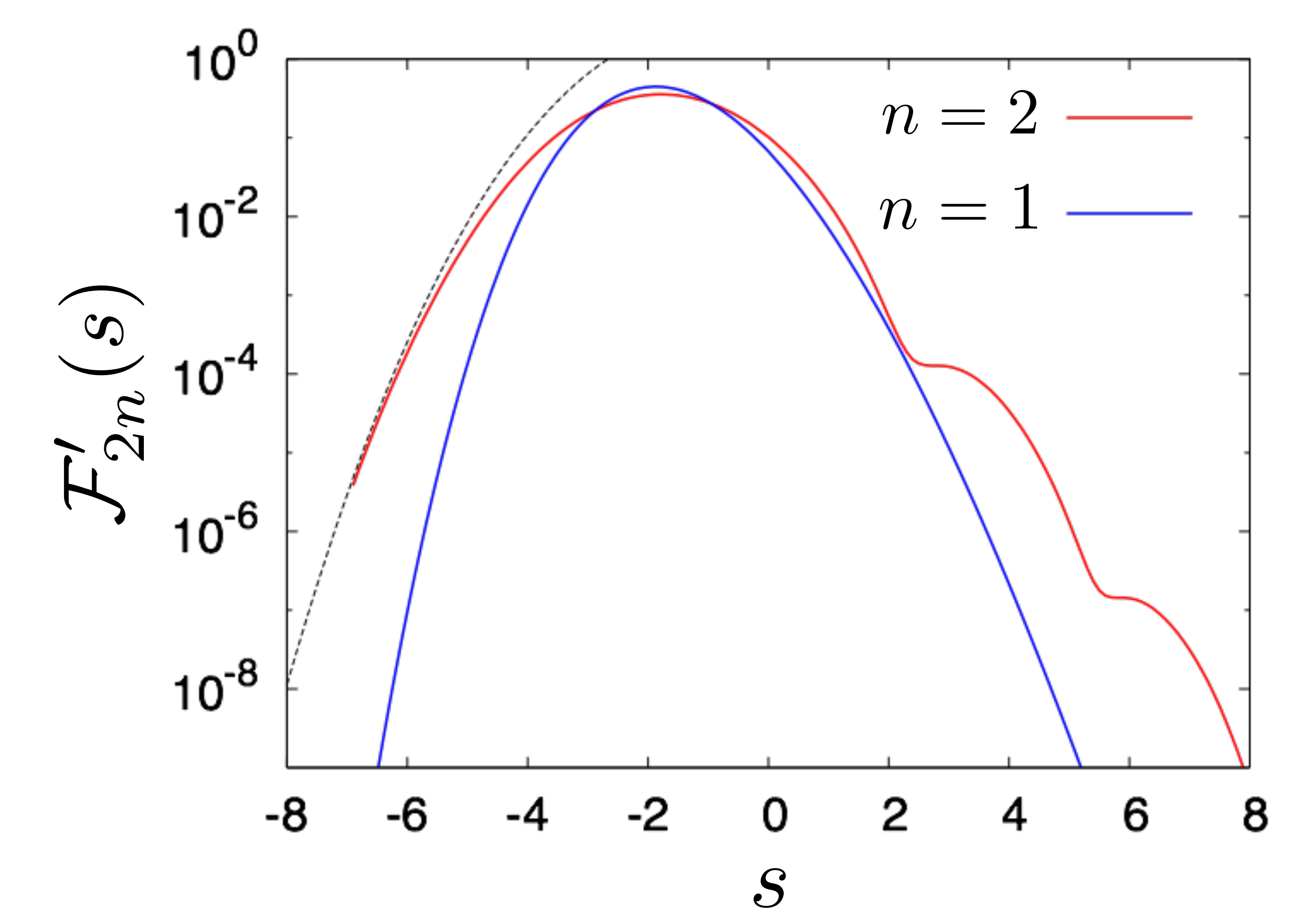}
\caption{Plot of the limiting PDF of $p_{\max}$, ${\cal F}'_{4}(s)$ (in red), corresponding to $n=2$, compared to the standard TW distribution ${\cal F}'_2(s)$ (in blue) corresponding to $n=1$. The dotted line corresponds to the large negative behaviour ${\cal F}'_{4}(s) \approx \exp (- \frac{\sqrt{8/3}}{15}\, s^{5/2})$, which is quite different from the TW distribution ${\cal F}_2'(s) \approx e^{-\frac{|s|^3}{12}}$ for $s \to -\infty$ [see Eq. (\ref{asympt_TW})]. Notice also the surprising oscillatory behavior of ${\cal F}'_4(s)$ for $s \to +\infty$ (see Ref. \cite{multicritical}) for mode details.}\label{Fig_F4}
\end{figure}

In \cite{multicritical} it was shown that the fluctuations of $p_{\max}$ 
are given by a set of new distributions, indexed by integer $n \geq 2$, which we called ${\cal F}_{2n}$, different from the usual Tracy-Widom distribution, which one gets for $n=1$ (see Fig. \ref{Fig_F4}). These are found to satisfy a hierarchy of Painlev\'e equations,
for different $n$, which generalize the one for the usual TW distribution. Such hierarchies are
also encountered in multicritical matrix models of interest in random surfaces and
string theory. This raises the possibility of interesting connections, yet to be explored.
We refer the reader to \cite{multicritical} for more details.

\section{Conclusion}\label{Sec:conclusion}

To conclude, we have shown how RMT techniques (in dimension $d=1$ and at temperature $T=0$), and more generally the methods borrowed from determinantal point processes provide the ideal tools to study in detail noninteracting trapped fermions. In particular, these methods turn out to be extremely useful to study the universal correlations that emerge at the edge of trapped Fermi gases, where the standard techniques, such as the Local Density Approximation (LDA), can not be applied. In $d=1$ and $T=0$ we have reviewed various classes of quantum potentials $V(x)$ that lead to
noninteracting Fermi systems which are in one-to-one correspondence with the standard unitary ensembles of RMT, namely GUE, JUE and LUE (see Fig. \ref{Fig_V_RMT}). In particular, the edge correlations are described by the Airy kernel (for smooth potentials leading to a ``soft edge'') and the Bessel kernel (for sufficiently singular potentials leading to a ``hard edge''). An interesting outcome of this mapping to RMT \cite{us_finiteT, fermions_review} is that the quantum fluctuations of the position of the rightmost fermion in a smooth potential (i.e. of the form $V(x) \sim |x|^p$ with $p>0$) are described at $T=0$, by the celebrated Tracy-Widom distribution \cite{TW94}. For noninteracting fermions in a quadratic potential, the mapping to GUE was also very useful to study the fluctuations of the number of particles inside an interval ${\cal I}$, in particular its variance \cite{marino_prl} which, for ${\cal I}$ inside the bulk, is related to the entanglement entropy of ${\cal I}$ with its complement~\cite{CLM15}. In this case, it is however quite difficult to study in detail the relation between the number variance and the entanglement entropy for a domain ${\cal I}$ close to the edge. Recently, it was shown that both quantities can actually be computed at the edge for a model of noninteracting fermions in a two-dimensional rotating harmonic trap \cite{LMS18}.

In higher dimensions $d \geq 2$, still at $T=0$, the connection to RMT is generically lost -- see however the case of $2d$-fermions trapped in a rotating harmonic potential mentioned above which can be mapped onto the so called Ginibre ensemble of RMT \cite{LMS18}. However, the universal correlations at the edge can still be studied using the tools of determinantal point processes. In this case, the associated kernels are given by generalisations of the Airy kernel (for smooth potentials) and of the Bessel kernel (for hard edge potentials), that depend non-trivially on the dimension $d$ \cite{DPMS:2015,fermions_review,LLMS17,LLMS18}. 

At finite temperature $T>0$, in the canonical ensemble where the number of fermions $N$ is fixed, the correlations are much harder to study, since the corresponding processes cease to be determinantal. Despite this, it is possible, in some cases, to obtain exact results for the linear statistics, i.e. for the distribution of physical observables of the form ${\cal O} = \sum_{i=1}^N f(x_i)$ of the positions $x_i$'s of the $N$ fermions~\cite{GMS17,GMST18}. To compute the correlations, one can use the equivalence between the thermodynamic ensembles, which is expected to hold for $N \gg 1$, and work in the grand-canonical ensemble. The great advantage of working in the grand-canonical ensemble is that the positions of the trapped fermions do, again, form a determinantal point processes. And in this case, one can also show that the fluctuations at the edge are governed by universal correlation kernels that depend on both the dimension $d \geq 1$ and the temperature $T > 0$, both for smooth \cite{us_finiteT,fermions_review} and hard-edge  \cite{LLMS17,LLMS18} potentials. In particular, in $d=1$ and $T>0$ (properly scaled with $N$, see Fig.~\ref{Fig_FiniteT_TW}), the distribution of the position of the rightmost fermion, in a smooth potential, is given by a finite temperature generalisation of the Tracy-Widom distribution \cite{us_finiteT, fermions_review, Joh07}. Interestingly, as noticed in \cite{us_finiteT}, the very same distribution [see Eq. (\ref{finite_T_Ai})] appears in the exact solution of the Kardar-Parisi-Zhang (KPZ) equation at finite time, and for droplet initial condition. Inspired by this connection between trapped fermions at finite temperature and the KPZ equation, further developments \cite{periodic_airy} have concerned the construction of a periodic version of the so called Airy process, which underlies the (spatial) fluctuations in the KPZ equation \cite{airy}. In particular, it was shown that this periodic Airy process describes the equilibrium (i.e. imaginary time) dynamics of trapped fermions near a (soft) edge. Recently, this periodic Airy process was found \cite{jeremie} to occur in combinatorics, in the context of the so called ``periodic Schur process'' \cite{borodin}. 

Finally, we have discussed the Wigner function, which is the natural observable to 
characterize the correlations in the phase space (i.e. in the space of position $x$ and momentum $p$). In particular it also exhibits an edge in the $(x,p)$-plane. The vicinity of this edge was studied, for smooth potentials, for any dimension $d \geq 1$, at $T=0$ as well as finite temperature $T>0$ and it was shown \cite{us_Wigner} that the Wigner function is described by a ``super-universal'', $d$-independent, scaling function at the edge [see Eq.~(\ref{scal2T})]. This work further lead to investigate the statistics of the {\it momenta} of trapped fermions, which, at $T=0$, also form a determinantal point process. In particular, in dimension $d=1$ and for anharmonic potential $V(x) \sim |x|^p$, with $p>0$, an interesting connection with multi-critical matrix models was unveiled \cite{multicritical}.

This set of results, obtained using the methods of RMT and determinantal point processes, raise open challenging questions. In particular, the calculations presented here concern noninteracting fermions. A natural and important question thus concerns the effects of interactions, both in the bulk and at the edge. This is particularly challenging in one dimension where the standard Fermi liquid theory fails \cite{giamarchi}. Similarly, one may wonder about the effects of quenched disorder. Finally, it would be very interesting to extend the methods presented here to non-equilibrium situations, for instance in the context of quantum quenches \cite{allegra,collura}. We hope that the methods and results reviewed here will stimulate further research along these lines.

\ack
We thank P. Calabrese, A. Grabsch, J. Grela, A. Krajenbrink, B. Lacroix-A-Chez-Toine, M. Mari\~no, R. Marino, C. Salomon, G. Salomon, C. Texier, P. Vivo and P.  Wiegmann for useful discussions. This research was partially supported by ANR grant ANR-17-CE30-0027-01 RaMaTraF.


\section*{References}


{}
\end{document}